\documentclass[aps,prc,amsmath,amssymb,reprint]{revtex4-1}

\usepackage{graphicx, fancybox}
\usepackage{amsmath,amssymb}
\usepackage[colorlinks=true, pdfstartview=FitV, linkcolor=red, citecolor=blue, urlcolor=blue]{hyperref}
\usepackage[usenames, dvipsnames]{color}
\usepackage{soul}
\usepackage{url}
\usepackage{appendix}
\usepackage{float}
\usepackage[normalem]{ulem}

\begin{document}

\title{The 3+1D initialization and evolution of the Glasma }

\author{Scott McDonald}
\thanks{Currently at Shift Energy Inc. 1 Germain St, Saint John, NB, E2L
4V1, Canada}
\affiliation{
Department of Physics, McGill University, 3600 University
Street, Montreal,
QC, H3A 2T8, Canada
}

\author{Sangyong Jeon}
 \affiliation{Department of Physics, McGill University, 3600 University
 Street, Montreal,
 QC, H3A 2T8, Canada}

 \author{Charles Gale}
 \affiliation{Department of Physics, McGill University, 3600 University
 Street, Montreal,
 QC, H3A 2T8, Canada}

\begin{abstract}

The IP-Glasma initial condition has been highly successful in the
phenomenology of ultra-relativistic heavy ion collisions. The assumption of
boost invariance, however, while good for collision energies probed at the
LHC, limits the use of IP-Glasma to the transverse dynamics of heavy ion
collision to near mid-rapidity. There is a wealth of physics to be explored
and understood in the longitudinal dynamics of heavy ion collisions, and a
full understanding of heavy ion collisions can only come from 3-dimensional
studies. In particular, long range rapidity correlations are seeded in the
initial collision and provide additional information on the high energy
nuclear wave functions that has thus far been inaccessible to the IP-Glasma
model. In this work, we introduce a way to extend the IP-Glasma model to
3+1-dimensions while preserving its key features.
\end{abstract}

\maketitle
\date{\today}

\section{Introduction}

Heavy ion collisions (HIC's) conducted at RHIC and the LHC are sufficiently
energetic to create a deconfined state of quarks and gluons known as Quark
Gluon Plasma (QGP). Due to their complexity and changing degrees of
freedom, no single model is able to describe the entirety of these
collisions, and thus they are modeled in stages, usually including
independent models for the initial state, a hydrodynamic (QGP) phase, and a
hadronic gas phase.

There is broad agreement in the field that nucleus-nucleus collisions
create QGP and that this exotic state of matter is governed by relativistic
fluid dynamics with an extremely small shear viscosity to entropy density
ratio (specific shear viscosity),
$\eta/s$ \cite{STAR:2005gfr,PHENIX:2004vcz,Arslandok:2023utm}.
It is similarly accepted that, as the fluid expands and
cools, it hadronizes and can be described by hadron gas dynamics
such as
those modeled with UrQMD \cite{Bass:1998ca} or SMASH \cite{Petersen:2018jag}.
The initial state, however, has not reached such a high level of consensus.

Because the outcome of hydrodynamic simulations are sensitive to the details of the
initial conditions, it is important to constrain the initial state before
strong statements can be made about details of the QGP phase, such
as the transport coefficients. In  order to do so, it is important to
explore both the transverse and the longitudinal dynamics of HIC's.
The transverse dynamics has been successfully explored by many via 2+1D models 
of the initial condition and the dynamics 
\cite{Sollfrank:1996hd,
Huovinen:1998tq,
Kolb:2000fha,
Gyulassy:2001kr,
Heinz:2005bw,
Huovinen:2006jp,
Baier:2006um,
Song:2007ux,
Shen:2010uy,
Shen:2011eg,
Qiu:2011hf,
Gale:2013da,
Jeon:2015dfa}.

The focus of this paper is on constructing an IP-Glasma based 3+1D initial
state model and exploring its physical consequences.
In order to do so, it is necessary to have 3+1D
simulations of HIC's. This has largely been achieved for the hydrodynamic
and hadronic phases of the QGP 
evolution \cite{Rischke:1995ir,Nonaka:2000ek,Hirano:2001yi,Schenke:2010rr,Schenke:2010nt}.
This paper sets out to generalize the
phenomenologically successful IP-Glasma model \cite{Schenke:2012fw,
Schenke:2012wb} to 3+1D so that the 3+1D machinery can be fully utilized.

IP-Glasma provides a 2+1D initial condition
that combines IP-Sat-inspired \cite{Kowalski:2003hm} small-$x$ gluon saturation
with classical Yang-Mills evolution. It has been extremely
successful in describing the transverse dynamics of heavy ion collisions
when used to initialize hydrodynamic simulations
\cite{Gale:2012rq,McDonald:2016vlt}.
This includes many
different observables across a wide range of collision systems and center
of mass energies.

The original IP-Glasma model assumes boost invariance, which simplifies the geometry
of heavy ion collisions to 2+1-dimensions. This allows for a direct analytic solutions of the
classical Yang-Mills field and also simplifies numerical evolution of the system.
This is a good approximation
near mid-rapidity at high energies such as those explored at the LHC, but
remains an approximation nonetheless. Furthermore, asymmetric systems such
as $p{+}A$ collisions cannot be accurately approximated as boost invariant due
to their large rapidity dependence.
This assumption also limits one to study only the transverse dynamics of heavy ion collisions. 
By relaxing boost invariance, one gains access to the longitudinal dynamics of
heavy ion collisions, where there is a wealth of physics to be explored and
understood.
In this work, we relax boost invariance
in the IP-Glasma framework by providing longitudinal structure
using the JIMWLK renormalization group equation
\cite{Ayala:1995hx,JalilianMarian:1996xn,Iancu:2001ad,
Blaizot:2002np,Gelis:2008rw,Gelis:2008ad,Lappi:2012vw}, 
and solve the Classical Yang-Mills (CYM) equations on a 3-dimensional 
lattice \cite{Romatschke:2005pm,Romatschke:2006nk,Fukushima:2006ax,Fukushima:2011nq,Berges:2012cj,Dusling:2012ig,
Epelbaum:2013waa,Gelfand:2016yho}.

The consequences of generalizing the IP-Glasma to 3+1D will be explored
through comparison with the boost invariant case. The 3+1D initial conditions
are evolved hydrodynamically using MUSIC 
\cite{Schenke:2010rr, Schenke:2010nt}.
As the system expands and cools, and hadronizes, UrQMD
is used to simulate resonance decays and hadronic
re-scatterings. Longitudinal observables are studied and compared to
experimental data.
One of the consequences we would like to study in this work is the longitudinal correlations.
Correlations and fluctuations present in the high energy nuclear wave
functions of the colliding nuclei are imprinted on the system during the
initial collision. Some of them will be preserved and detected in the final state and some will not,
depending on their nature and strength as well as those of the subsequent
evolution of the fireball. It is clear, however, that correlations present
at the initial collision constitute the upper bound for long range
correlations in rapidity.

There are many reasons why we would like to explore longitudinal dynamics of relativistic heavy ion collisions. 
One important reason is to see whether our understanding of the QGP dynamics mostly gained from 2+1D
studies will still hold in describing the longitudinal dynamics. For instance, we would like to see whether
the values of the viscosities extracted in 3+1D study are consistent with those extracted in 2+1D study.
We would also like to investigate to what extent the boost-invariant approximation
breaks in the plateau region around the mid-rapidity and what breaks it.
It will be also interesting (although we leave it for future study) to see how the asymmetry in the size of the colliding nuclei affects
the longitudinal dynamics.
Other important topics include the effect of 3+1D evolution to the longitudinal flux tube and 
the classical gluon field's influence on the propagation of jet partons inside and outside of the plateau region.

In the following, we first briefly describe the 2+1D IP-Glasma initial conditions in section 
\ref{sec:Init_Cond_2}. Generalization to 3+1D is explained in sections 
\ref{sec:Gen_to_3D} through \ref{Section:JIMWLK}.
The differences between the 2+1D evolutions and the 3+1D evolutions are 
highlighted in sections \ref{sec:Pressure} and \ref{sec:2DLimit}.
In section \ref{sec:results}, the 3+1D results are compared with ALICE data and we conclude in 
section \ref{sec:conclusion}.

\section{Initial Conditions in 2+1D}
\label{sec:Init_Cond_2}

The large occupation number of small-$x$ gluons at early times in HIC's
means that they can be treated, to good approximation, as classical fields.
The relevant equations of motion are then the Classical Yang-Mills (CYM)
equations,
\begin{equation}
    [D_\mu, F^{\mu\nu}] = J^{\nu}.
\end{equation}
In this paper, we use the convention $D_\mu = \partial_\mu - igA_\mu$ and the mostly negative metric.
Under the assumption that the source particles are moving with the speed of
light in the same direction, the source terms,
comprised of the large momentum fraction (large-$x$) 
partons in the individual nuclei,
propagate undeflected on the light-cone,
\begin{equation}
    J^{\nu}=\delta^{\nu \pm}\rho_{A(B)}(x^\mp, \mathbf{x}).
\end{equation}
where $x^{\pm} = (t\pm z)/\sqrt{2}$ are the light-cone coordinates.
The upper signs are for the projectile nucleus $A$ moving in the positive $z$
direction and the lower signs are for the target nucleus $B$ moving in the
negative $z$ direction.
In this limit, it is possible to derive an analytic solution to the initial
gauge fields immediately following the collision in terms of the gauge
fields of the pre-collision nuclei $A$ and $B$.

In light-cone coordinates and Lorentz gauge, the pre-collision CYM equations reduce to
the Poisson equation
\begin{equation}
\label{eq:MV_Sol}
    A^{\pm}_{A(B)} = -\frac{\rho_{A(B)}}{\nabla_\perp^2 - m^2}.
\end{equation}
where $m = 0.4\,\hbox{GeV}$ is an infrared regulator that models
the colour neutrality scale.
Here $A^+$ is for the projectile nucleus and $A^-$ is for the target nucleus.
These gauge fields can be gauge-transformed
to the light-cone gauge 
by using the following path-ordered Wilson lines
\begin{equation}
    V_{A(B)}({\bf x}_\perp)=P {\rm exp}\Big(-ig\int dx^{\mp}
\frac{\rho_{A(B)}(x^{\mp}, x_\perp)}{\nabla_\perp^2-m^2}\Big)
\label{eq:V_AB}
\end{equation}
whose the discretized form can be written as \cite{Lappi:2007ku},
\begin{equation}\label{Eq:Wilson_line}
V_{A,B}(\mathbf{x_\perp})=\prod_{i=1}^{N^\mp}
\exp\Big({-ig\frac{\rho_i^{A,B}(x_\perp)}{\nabla_\perp^2-m^2}}\Big)
\end{equation}
where $N^\mp$ is typically set to $50$.
The pre-collision
gauge fields then become purely transverse
\begin{align}
\label{eq:MV_Sol_2}
    A_i &= \frac{i}{g}V \partial_i V^\dagger
\\
\label{eq:Aetazero}
    A^{\pm} &=0.
\end{align}
This is the celebrated McLerran-Venugopalan (MV) model 
\cite{McLerran:1993ni,McLerran:1993ka}.

\begin{figure*}
    \centering
    \begin{minipage}{.5\textwidth}
    \includegraphics[width=1.\linewidth]{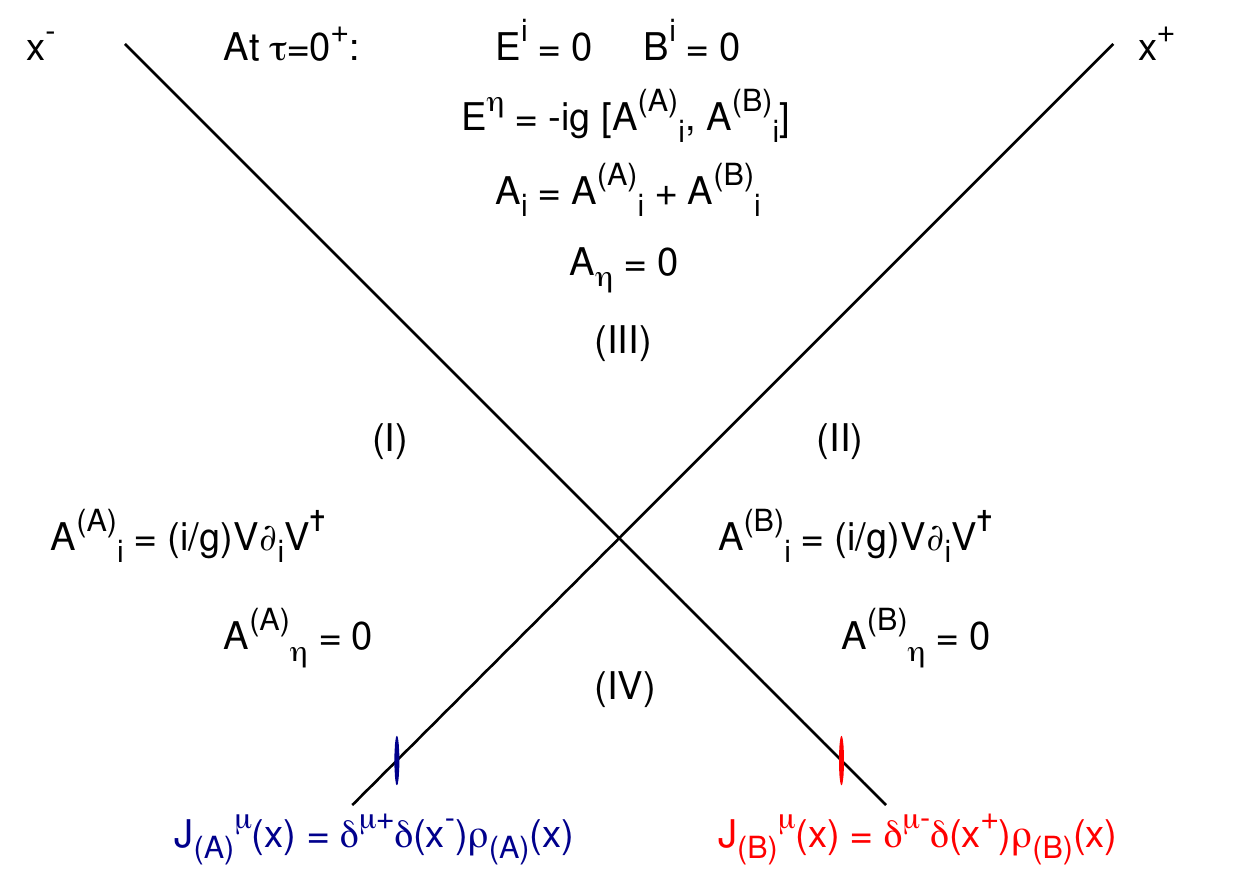}
    \caption{A summary of the 2+1D initial conditions}
    \label{fig:2D_init_summary}
    \end{minipage}%
\begin{minipage}{0.5\textwidth}
    \includegraphics[width=1.\textwidth]{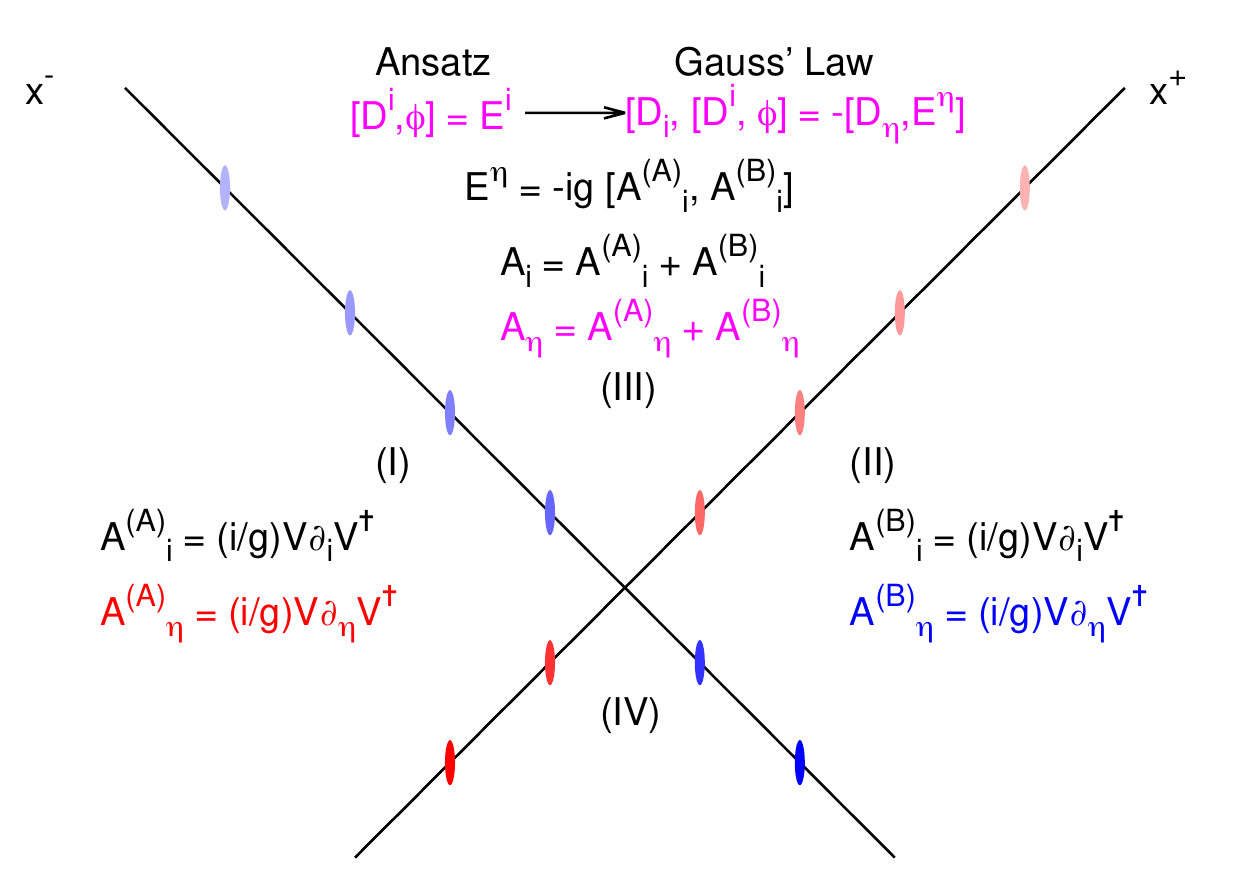}
    \caption{A summary of the 3+1D initial conditions.}
    \label{fig:3D_init_summary}
    \end{minipage}
\end{figure*}

These pre-collision fields can be related to the post-collision gauge
fields that reside in the forward light cone by matching the fields on the
light-cone boundary, including the source terms. 
This matching yields the initial Glasma fields \cite{Kovner:1995ts,Kovner:1995ja}
given by
\begin{align}
A_0^{i}&=A_{A}^{i}+A_{B}^{i},
\label{eq:A_perp}
\\
E_0^{\eta}&=-{ig}[A_{A}^{i},A_{B}^{i}],
\label{eq:E_eta}
\end{align}
where again the subscripts $A$ and $B$ refer to the projectile nucleus
and the target nucleus, respectively.
The coordinate system for the Glasma field is the Milne 
coordinate system where $\tau = \sqrt{t^2 - z^2}$ and
$\eta = \tanh^{-1}(z/t)$. The gauge condition for the Glasma fields is $A^\tau = 0$.
In 2+1D, one can identify $E^\eta$ with $-2A^\eta$ using
\begin{align}
E^\eta &= {1\over \tau}\partial_\tau A_\eta 
= -{1\over \tau}\partial_\tau (\tau^2 A^\eta)
\nonumber\\
&= -2A^\eta -\tau \partial_\tau A^\eta
\label{eq:E_eta_A_eta}
\end{align}
where we used the fact that $A_\eta = -\tau^2 A^\eta$ 
and assumed that the second term vanishes as $\tau\to 0$.
For a visual summary of the 2+1D initial condition, see 
Fig.~\ref{fig:2D_init_summary}.

Having determined the initial gauge fields and the longitudinal electric
field, it remains to specify the transverse electric field, which must
satisfy Gauss' Law,
\begin{equation}\label{eq:gauss}
    [D_\eta, E_0^\eta]+[D_i,E_0^i]=0.
\end{equation}
The boost invariance of the system make Gauss' Law trivial, due to
vanishing derivatives in $\eta$. The resulting solution is simply $E^i_0=0$.
This solution is not unique.
Any vector field $e_0^i$ that satisfies $[D_i, e_0^i] = 0$ can be a solution.
However, $E_0^i = 0$ is the most natural choice
in view of the fact that the initial transverse chromo-magnetic fields $B_0^i = 0$
because the system is boost-invariant 
and we assumed $A_\eta = 0$ at $\tau = 0^+$.

For non-zero gradients in the rapidity-direction,
$E_0^i$ and $B_0^i$
become non-zero, and their magnitudes are determined by the size of the
$\eta$ gradients. These, in turn, come from the rapidity dependence of the
model. In the case of the current work, the rapidity dependence comes from
the JIMWLK renormalization group equation, to be discussed in section
\ref{Section:JIMWLK}.

\section{Generalizing to 3+1D}
\label{sec:Gen_to_3D}

The beauty of 2+1D IP-Glasma formulation is the availability of 
the exact solution of the classical Yang-Mill's equation Eq.(\ref{eq:MV_Sol}) 
in the infinite momentum (equivalently, in the infinite beam rapidity) limit.
Once this condition is relaxed, exact solutions are no longer available.
Some possibilities to resolve this issue include:
One can try to solve the classical Yang-Mills equations
numerically provided that the source profile of each nucleus at finite velocity
is known.
Or one can make Abelian assumption to solve the 3+1D CYM equations analytically
as in Refs.\cite{Lam:2000nz,Ozonder:2013moa}.
One can also try to modify the 2+1D solution in such a way to approximate
the physical situation. 
So far, to the authors' best knowledge,
most attempts at generalizing the MV model and IP-Glasma model
fall into the last category
\cite{McLerran:1997fk,
Gelfand:2016yho,
Schenke:2016ksl,
Schlichting:2020wrv,
Ipp:2021lwz,
Schenke:2022mjv,
Ipp:2022lid}
and it is also the route we will take.
(Other non IP-Glasma-related approaches such as flux-tube/string type models 
also exist \cite{Werner:2010aa,Shen:2017fnn}.)

Our general strategy is somewhat similar to the one employed in Ref.\cite{Schenke:2016ksl}
but not exactly the same.
Consider the usual MV solution
of the 2+1D Yang-Mills equation Eq.(\ref{eq:MV_Sol}) (equivalently, Eq.(\ref{eq:MV_Sol_2}))
given the colour charge density $\rho_{A(B)}$.
Since the colour charge density $\rho_{A(B)}$ does not depend on the rapidity, neither does the gluon
field $A^i_{A(B)}$. In other words, in any boosted frame, $A^i_{A(B)}$ looks exactly the same, and
hence, the resulting glasma field $A^i = A_A^i + A_B^i$ in any boosted frame looks exactly the same, 
too. This implies that the produced Glasma is boost-invariant.

\begin{figure*}
    \centering
    \includegraphics[width=0.8\textwidth]{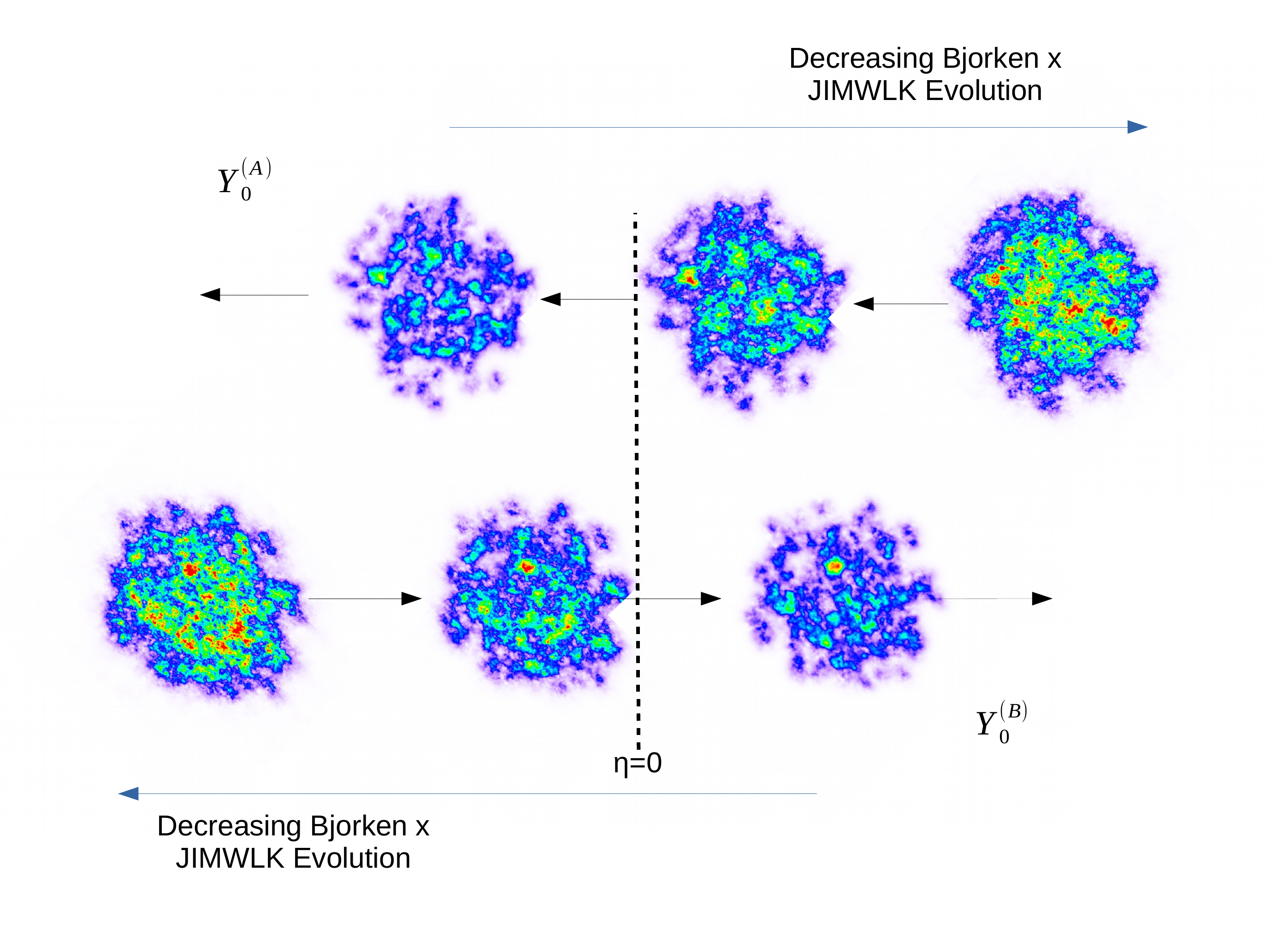}
    \caption{Two nuclei evolving in rapidity via the JIMWLK equations.  At
$Y^{(0)}_A$ and $Y^{(0)}_B$ the Wilson lines are determined via equation
(\ref{Eq:Wilson_line}). Then these Wilson Lines are evolved via equation
(\ref{eq:LangevinStep}). Plotted are snapshots of the quantity
$\frac{1}{N_c}{\rm Tr}(V-1)$, a proxy for gluon density. It is possible to
see as the JIMWLK evolution proceeds to smaller Bjorken-$x$, the gluon
density increases while the large scale geometry of the nuclear structure
persists.}
    \label{fig:boost_jimwlk}
\end{figure*}

Once we add quantum fluctuations, however,
the JIMWLK evolutions 
of the colour densities
break the boost invariance by introducing a reference rapidity.
In this way, the gluon densities of the projectile and the target nuclei can look {\em different} in
different boosted frame, equivalently at different space-time rapidity
$\eta = \tanh^{-1}(t/z)$. 
Consider a world-line passing through the origin of the center of mass frame so that
$z/t = v_z$ is a constant. This also represents a world-line where the space-time rapidity 
$\eta$ is constant.
One can therefore get the initial condition at $\eta$ by considering how the projectile and
the target nuclei appear in the frame boosted by $v_z = \tanh\eta$.
This is illustrated in Fig.~\ref{fig:boost_jimwlk}.

In Ref.\cite{Schenke:2016ksl}, this idea was used to get the initial condition for the 3+1D
evolution of the glasma field.
The difference here
is in the way the longitudinal dynamics is treated. In Ref.\cite{Schenke:2016ksl}, each
transverse plane at different $\eta$ evolves independent of each other 
following the usual 2+1D IP-Glasma formulation of the initial condition and evolution.
On the other hand, we keep the longitudinal interaction between the transverse planes 
at different rapidities. To do so, however, complicates not only the evolution of the system but also
the initial condition.

\section{Initial $A_\eta$ and $E_i$}\label{sec:3D_init_gauge_field}
\label{sec:Transverse_fields}

There has been significant effort in recent years on 3+1D Classical Yang
Mills in the context of heavy ion collisions.  Each of these efforts has
implemented some type of rapidity dependence, whether it be through
rapidity fluctuations 
\cite{Romatschke:2005pm,Romatschke:2006nk,Epelbaum:2014xea}, 
JIMWLK
evolution \cite{Schenke:2016ksl}, or 
colour sources \cite{Gelfand:2016yho,Schlichting:2020wrv,Ipp:2021lwz}.
In this work, we extend the initial
conditions themselves to be able to accommodate a non boost-invariant
setup.

\begin{figure}[h!]
    \centering
    \includegraphics[scale=0.3]{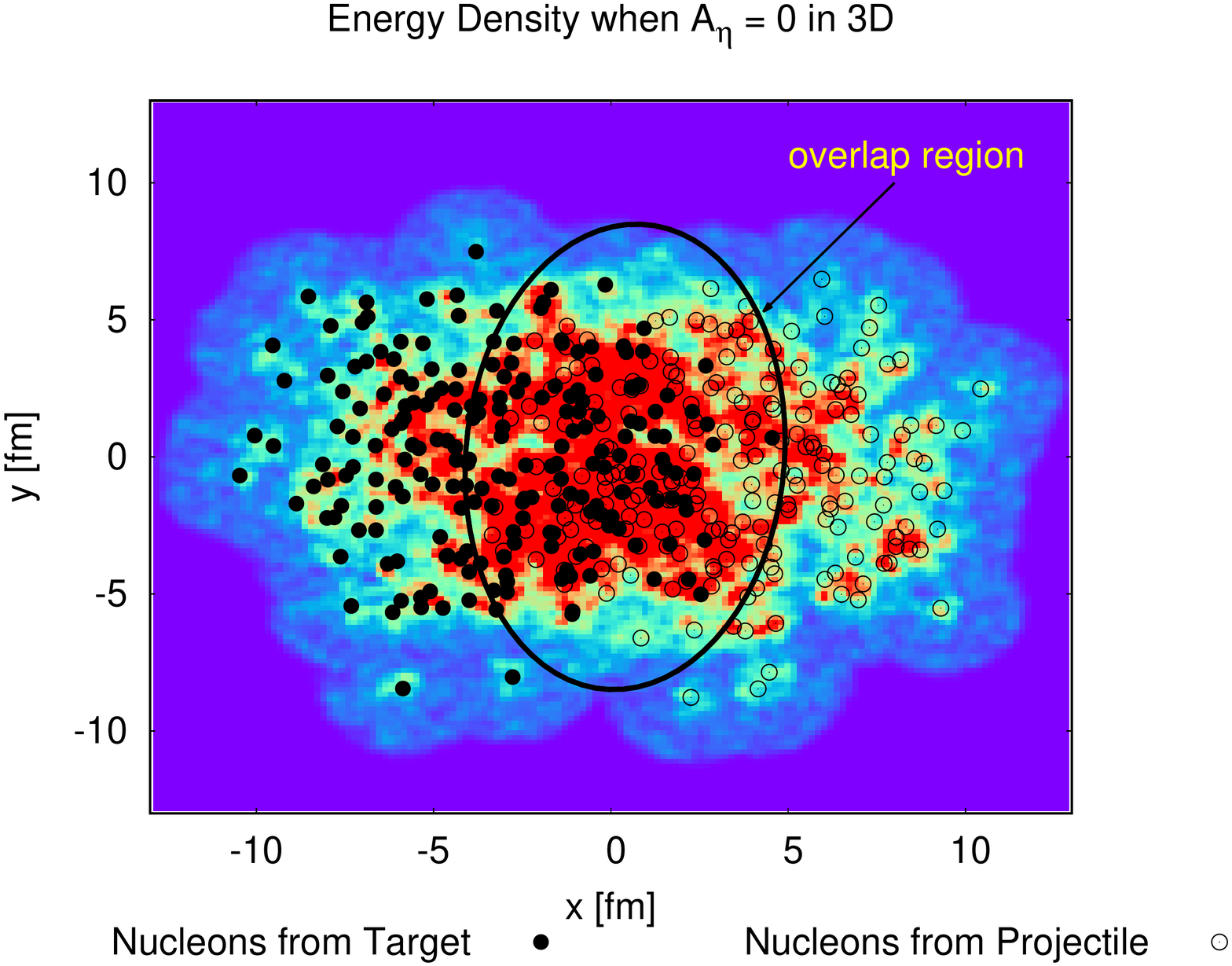}

    \includegraphics[scale=0.3]{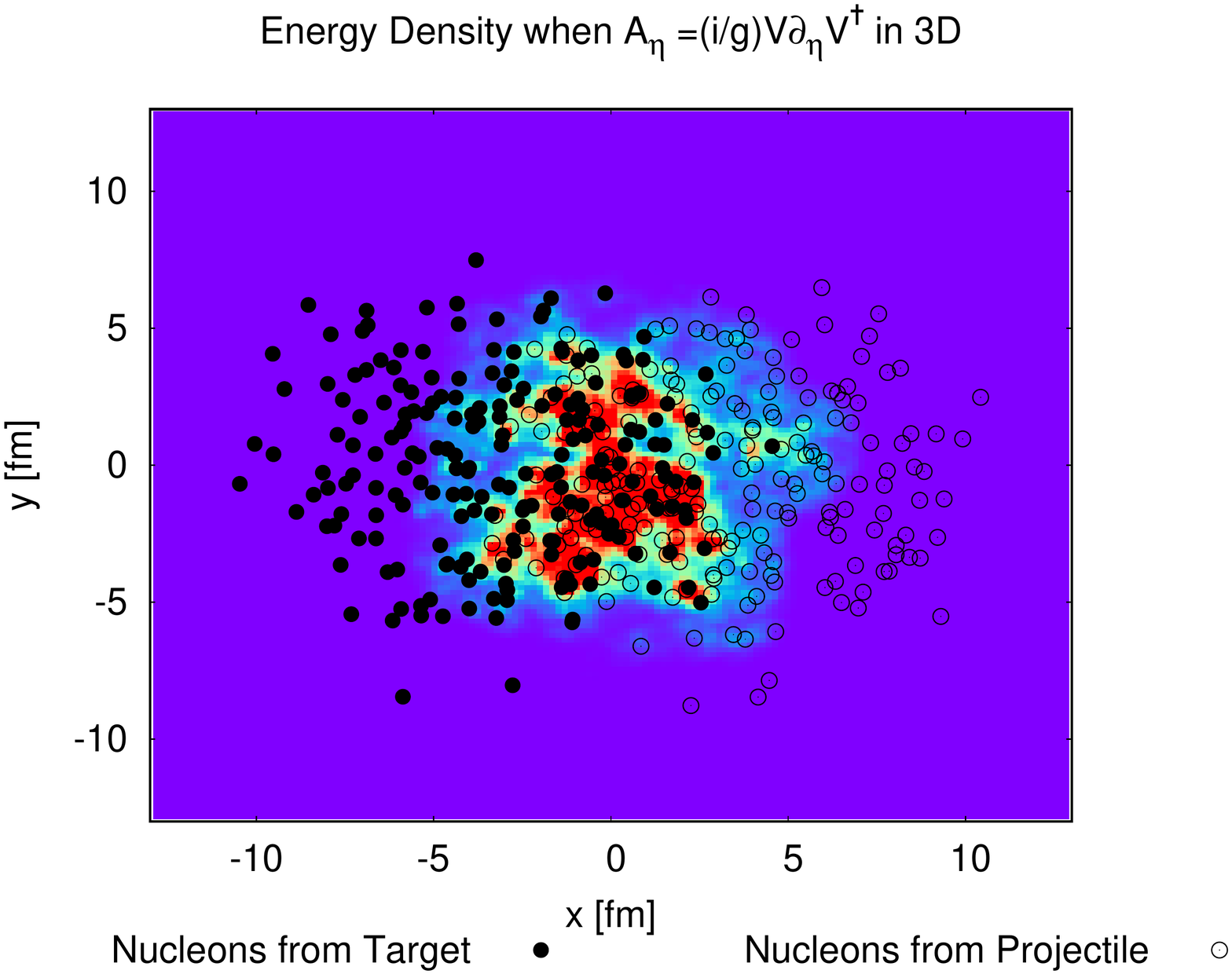}
    \caption{The upper figure demonstrates the problem of initializing
$A_\eta=0$ on 3-dimensional lattice, namely that there is non-zero energy density
outside of the overlap region of the two nuclei. The lower panel, in
contrast, shows that initializing according to Eq.(\ref{fig:Aeta_nonzero})
eliminates this problem. }
    \label{fig:Aeta_nonzero}
\end{figure}

One of the consequences of having an $\eta$ dependence is that the usual 2+1D
solution, $A_i = (i/g)V\partial_i A^\dagger, A_\eta = 0$, is no longer pure-gauge in space.
This introduces a problem in energy deposition because the field strength $F_{\eta i}$ no longer
vanishes outside the overlap region. 
This chromo-magnetic field component automatically vanishes in the 2+1D case outside the overlap.
However, in 3-dimensions, the derivative in $\eta$ no longer vanishes and hence
if one were to use the 2+1D MV solution for individual
nuclei, one would find
$ F_{\eta i} = \partial_{\eta}A_i \ne 0$.
This means that 
non-zero energy density would appear
in the transverse plane wherever a single nucleus had
non-zero gauge field, rather than solely in the interaction region. This
phenomenon can be seen clearly in Fig.~\ref{fig:Aeta_nonzero}.

This undesirable feature can conceivably be dealt with in two ways. 
One is to just remove the energy density from the
positions where either $A_A^i$ or $A_B^i$ vanishes. This option, however, is 
ambiguous since it is not clear whether any subtraction should be made in the 
regions where neither of
the two fields vanishes. Another more natural option is 
to generalize the initial condition by modifying the longitudinal gluon fields as
\begin{align}
A^{A\,(B)}_{0,\eta} &= \frac{i}{g}V_A \partial_\eta V_B^\dagger
\label{eq:A_eta_init}
\\
A_{0,\eta} &= A^{A}_\eta + A^B_\eta
\label{eq:A_eta_init_sum}
\end{align}
This has the advantage of retaining the feature that each individual
nucleus remains pure gauge in space, while reducing to the boost invariant case 
where derivatives in $\eta$ vanish. 
Recall that the field strength tensor of a pure gauge vanishes, and
thus does not contribute to the energy density. For a visual summary of the
3+1D initial condition, see Fig.~\ref{fig:3D_init_summary}.

It is worth noting here that 
in general, initial conditions are needed for the dynamic variable
$A_\eta$ and its conjugate momentum $E^\eta$ but not $A^\eta$.
The fact that we could specify the initial value for
$A^\eta$ as $-E_0^\eta/2$
in 2+1D is an artifact of assuming that the $-\tau\partial_\tau A^\eta$ term 
in Eq.(\ref{eq:E_eta_A_eta}) vanishes in the $\tau\to 0^+$ limit.
Such an assumption forces the behaviour of $A_\eta$ 
in the small $\tau$ limit to be
$A_{\eta} = (E_0^\eta/2)\tau^2 + O(\tau^3)$ and 
hence forces the initial $A_\eta$ to vanish.
However, this is not the only possibility. 
One can have
\begin{equation}
A_\eta(\tau) = 
A_{0,\eta}
+{E_0^\eta\over 2} \tau^2
+ O(\tau^3)
\end{equation}
and still get
\begin{equation}
\lim_{\tau\to 0^+} E^\eta
=
\lim_{\tau\to 0^+}\left({1\over \tau}\partial_\tau A_\eta\right) 
=
E_0^\eta
\end{equation}
as long as both $A_{0,\eta}$ and $E_0^\eta$ depend
only on ${\bf x}_\perp$ and $\eta$ and not on $\tau$. 
Hence, in the absence of any additional conditions,
$A_{0,\eta}$ is quite arbitrary.
In 2+1D, it is convenient to choose $A_{0,\eta} = 0$.
In 3+1D, we can exploit this freedom to consistently remove unwanted energy deposits.

Another consequence of having the $\eta$ dependence
is that the solution to Gauss' Law is now non-trivial. 
In fact, Gauss' law is under-constrained,
as it provides only one equation for two unknown fields, $E^x$ and $E^y$. It
is possible to find a solution by relating the two unknown fields through
the following ansatz
\begin{equation}\label{eq:phi_ansatz}
    E_0^i= [D^i,\phi_0]
\end{equation}
This ansatz turns Gauss' Law into the covariant Poisson equation
\begin{equation}
    [D_i,[D^i,\phi_0]] = -[D_\eta, E_0^\eta]
\end{equation}
which can be solved iteratively through a modified Jacobi method (see
Appendix \ref{appendix:gauss_law_numerical} for numerical details). This
ansatz leads to a solution to Gauss' Law in the non-boost invariant system.
This solution is, however, not unique. One can always
add  another vector field $e_0^i$ that 
is divergenceless in 2D ($[D_i, e_0^i] = 0$) 
and still satisfy the Gauss law.
In this study, we simply set $e_0^i = 0$ 
which is consistent with the conditions that the initial Glasma field should vanish outside
the interaction region and that the total energy deposit 
should have a reasonable value for RHIC and the LHC heavy ion collisions.
In our simulations, the lattice equation of motion preserves the lattice Gauss' law.

\section{JIMWLK Evolution}
\label{Section:JIMWLK}

The Color Glass Condensate (CGC) is predicated on the idea that the gluon
density of high energy nuclei will begin to saturate as the gluon density
becomes sufficiently high for gluon recombination to compete with gluon
radiation. It relies on a separation of scales, in which the large momentum
fraction, or large-$x$, partons serve as sources for the small-$x$ gluons.

The JIMWLK (Jalilian-Marian, Iancu, McLerran, Weigert, Leonidov, Kovner)
\cite{JalilianMarian:1996xn,Iancu:2001ad} 
renormalization group equation integrates out
the quantum fluctuations around the classical background field and change
the effective source term for the small-$x$ gluons. In this way, the JIMWLK
evolution introduces a rapidity dependent charge per unit area,
while preserving the form of the gluon Lagrangian. 
The JIMWLK evolution gives the model
its rapidity dependence through the stochastic gluon radiation that follows
from the rapidity evolution.

The form of the JIMWLK equation used in this work is from Ref.\cite{Lappi:2012vw} and given
in terms of the Langevin step,
\begin{equation}\label{eq:LangevinStep}
\begin{split}
     V_{A,B}(\mathbf{x}, Y+dY)&=\exp\left({-i\frac{\sqrt{
dY}}{\pi}\int_{\mathbf{u}} \mathbf{K}_{\mathbf{x}-\mathbf{u}} 
\cdot( V_u \boldsymbol{\zeta}_u V_u^\dagger)}\right)\\
     & V_{A,B}(\mathbf{x}, Y) \exp
\left({i\frac{\sqrt{dY}}{\pi}\int_{\mathbf{v}}
\mathbf{K}_{\mathbf{x}-\mathbf{v}} \cdot
\mathbf{\boldsymbol{\zeta}}_v}\right)
\end{split}
\end{equation}
 where $\boldsymbol{\zeta}_z=\left\{ \zeta_1^a(\mathbf{z}, Y)
t^a,\zeta_2^a(\mathbf{z}, Y) t^a \right\}$ is a random variable
and $V_u = V_{A,B}({\bf u}, Y)$. Here $Y$ can be either the dynamic rapidity
or the space-time rapidity.
The correlator for $\boldsymbol{\zeta}$ in this case is given by,
\begin{equation}\label{eq:zeta_zeta_corr}
\langle \mathbf{\zeta}^{a,i}(\mathbf{x}, Y_1)\mathbf{\zeta}^{b,j}(\mathbf{y},
Y_2) \rangle=\delta^{ab}\delta^{ij}\delta^{Y_1 Y_2}\int
\frac{d^2\mathbf{k}}{(2\pi)^2}e^{i\mathbf{k} \cdot (\mathbf{x-y})}
\alpha_s(\mathbf{k}).
  \end{equation}
where the noise correlator has a Kronecker delta for $(Y_1,Y_2)$, rather than
a delta function, because the $1/dY$ has already been incorporated into
Eq.(\ref{eq:LangevinStep}). The modified kernel, as used in
Ref.\cite{Schlichting:2014ipa}, is given by,
\begin{align}
{\bf K}_{\mathbf{x-z}}&=m|\mathbf{x}-\mathbf{z}|K_1(m|\mathbf{x}-\mathbf{z}|)
\frac{\mathbf{(x-z)}}{|{\bf x}-{\bf z}|^2}
\end{align}
where $K_1(x)$ is the Bessel function of the second kind.

The expression in the exponent of Eq.~(\ref{eq:LangevinStep}) is computed
by Fourier transforming the kernel and the noise terms, thus turning the
2-dimensional integration into a convolution \cite{Rummukainen:2003ns}.
This improves numerical speed considerably. The Fourier transform of the
kernel is given by
\begin{equation}
    {\bf K}_{\mathbf{k}} = \frac{2\pi i \mathbf{k}}{k^2+m^2}.
\end{equation}
The form of the running coupling is taken to be
\begin{equation}
    \alpha_s(\mathbf{k}) = \frac{4\pi}{\beta
\ln{[(\frac{\mu_0^2}{\Lambda_{QCD}^2})^{1/c}+(\frac{\mathbf{k}^2}{\Lambda_{QCD}^2})^{1/c}]^c}}
\label{eq:alpha_s}
\end{equation}
with $\beta =11 - 2N_f/3$, $\Lambda_{QCD}=0.2$ GeV, $c=0.2$, and
$\mu_0=0.4$ GeV following the prescription in Ref.\cite{Lappi:2012vw}.

In principle, the scale at which the noise fluctuations occur should not
exceed the saturation scale, as it is the only physical scale in the
problem. However, the noise correlator is a 3-dimensional delta-function,
which means the numerical fluctuations take place at the scale of the
inverse lattice spacing $\approx 1/a$. Incorporating the running coupling
in the kernel acts to filter out higher $|{\bf k}|$ modes. Physically, this
means that the scale of the running coupling is taken to be that of the
emitted gluon.

\section{Equations of Motion}
\label{sec:EoM}

\begin{figure}[th]
  \centering
        \includegraphics[width=0.35\textwidth]{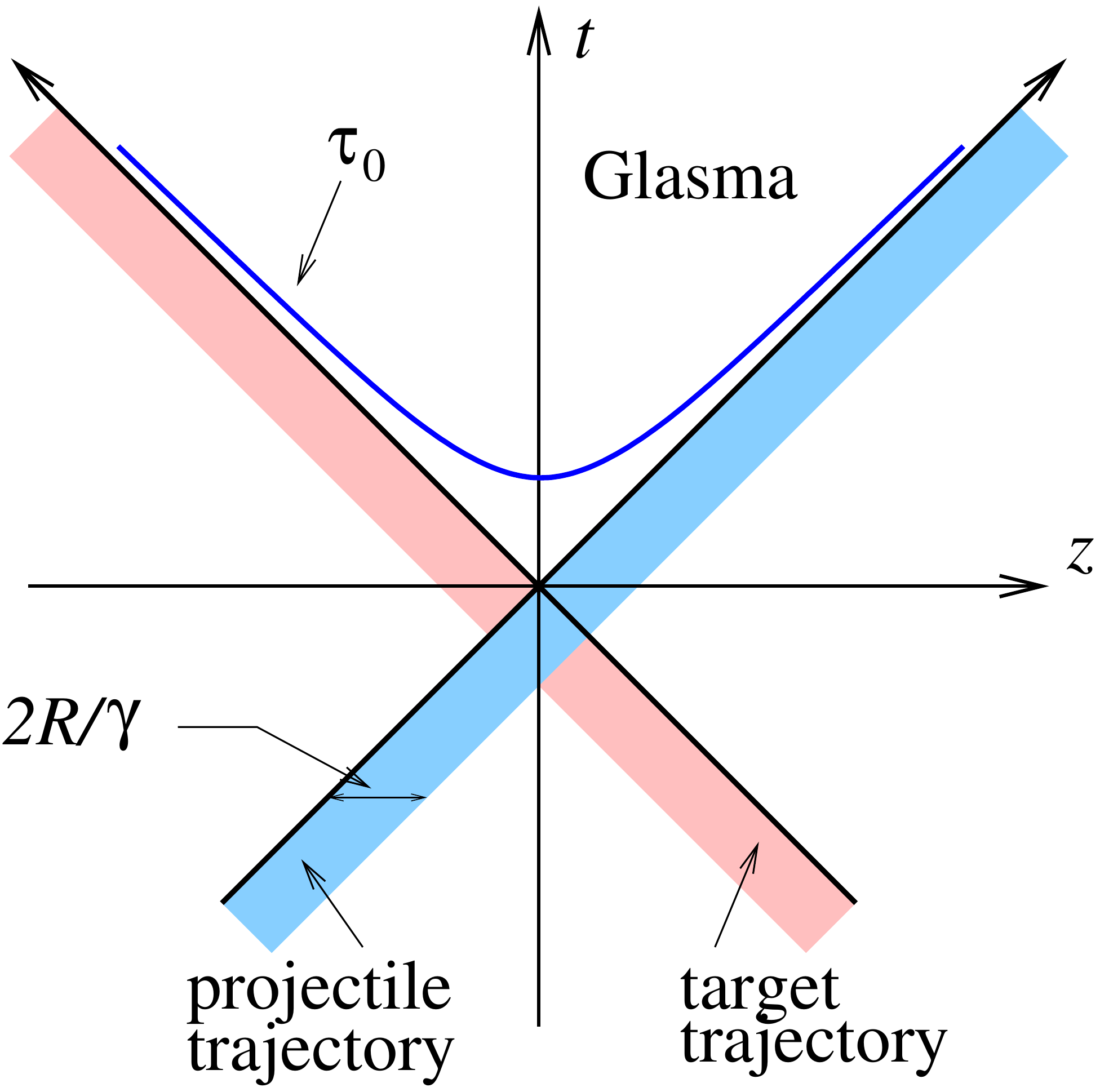}
\caption{Positioning of the $t$ and $z$ axes.}
\label{fig:finite_thickness}
\end{figure}

The evolution of Glasma in this study is performed in the $\tau{-}\eta$ coordinate system.
As in 2+1D, the source terms in
the Lagrangian are assumed to be eikonal and propagate along the light-cone
axes. Furthermore we set up our coordinate system in such a way that the source
currents are below the light-cone axes
as shown in Fig.~\ref{fig:finite_thickness}.
In this way, even though the sources have a finite thickness, the forward light-cone is
source-free and the equation of motion is simply
\begin{equation}
    [D_\mu, F^{\mu \nu}] = 0
\end{equation}
for $\tau > 0$.
Had we set up our coordinate system so that the leading edges of the nuclei define
$x^\pm$ instead of the trailing edges, the forward light-cone would not be source-free. 
Had we used Minkowski coordinate system, there will also be regions in $z$
that are not source-free \cite{Gelfand:2016yho,Schlichting:2020wrv}. 

The degrees of freedom that are evolved explicitly in time are the electric
field and the gauge links in the temporal gauge ($A^\tau=0$). 
Starting from the Hamiltonian
\begin{equation}
H = \tau\int d\eta \int d^2 x_\perp
\left( \epsilon_\eta + \epsilon_x + \epsilon_y\right)
\end{equation}
where
\begin{align}
     \epsilon_{i=x, y} &= \frac{1}{2}\frac{1}{\tau^2}\Big[(E^i)^2+(B^i)^2\Big]
     \label{eq:epsilon_i}
     \\
     \epsilon_\eta &= \frac{1}{2}\Big [(E^\eta)^2+(B^{\eta})^2\Big].
     \label{eq:epsilon_eta}
\end{align}
are the transverse and longitudinal energy densities,
the Hamiltonian equations of motion for the gauge fields can be derived as
\begin{align}
    \tau \partial_\tau A_i &= E^i
    \\
    \frac{1}{\tau} \partial_\tau A_\eta &= E^\eta
\end{align}
and
\begin{align}
 \partial_\tau E^i &= \frac{1}{\tau}\big[D_\eta, F_{\eta i}\big]+\tau\big[D_j, F_{ji}\big]
 \\
 \partial_\tau E^\eta &= \frac{1}{\tau}\big[D_j, F_{j \eta}\big]
\end{align}
The lattice version of these equations and the numerical method we use to solve them
closely follow those in Ref.\cite{Romatschke:2006nk}.

Because of $1/\tau$ factors in the equations,
the initial time cannot really be pushed to $\tau = 0$.
At LHC energies, the saturation scale used in this study is $Q_s \approx 2 - 5\,{\rm GeV}$
which corresponds to $1/Q_s \approx 0.04 - 0.2\,{\rm fm}$.
In Ref.\cite{Romatschke:2006nk}, it was argued that the initial proper time
$\tau_{0}$ should be much smaller than
$1/Q_s$. In this work, the initial time is set to $\tau_{0} = 0.01 {\rm \, fm}$.

To deal with the $1/\tau$ factors in the equations,
early time evolutions require very small time steps so that $\Delta\tau/\tau \ll 1$. 
For this reason, variable time steps are employed in the following form,
\begin{equation}
    \Delta\tau = \Delta\xi \tanh{\frac{\tau}{T_0}}.
\end{equation}
or $\xi = T_0\ln(\sinh(\tau/T_0))$.
This form interpolates between two limiting behaviours.  For $\tau \ll T_0$, the time step
behaves like $\Delta\tau \approx \Delta\xi (\frac{\tau}{T_0})$ while for $\tau>T_0$ it behaves like 
$\Delta\tau \approx \Delta\xi$ with $\Delta\xi$ fixed. This achieves the goal of producing small
time steps for small $\tau$ and larger equal time steps for later times when
the $1/\tau$ factors are no longer very large. In this work, $T_0$ is set to $0.2\,{\rm fm}$
and $\Delta\xi$ is set to $\tau_0/2 = 0.005\,{\rm fm}$.

\section{Fields and Pressure}
\label{sec:Pressure}

As already discussed, the initial transverse chromo-electric and
chromo-magnetic fields vanish in the boost invariant case. In 3+1D, this is
no longer the case and the transverse fields actually dominate the energy
density at early times due to the factor of $1/\tau^2$ in their
contribution to the energy density.

\begin{figure}[hb!]
  \centering
        \includegraphics[width=0.5\textwidth]{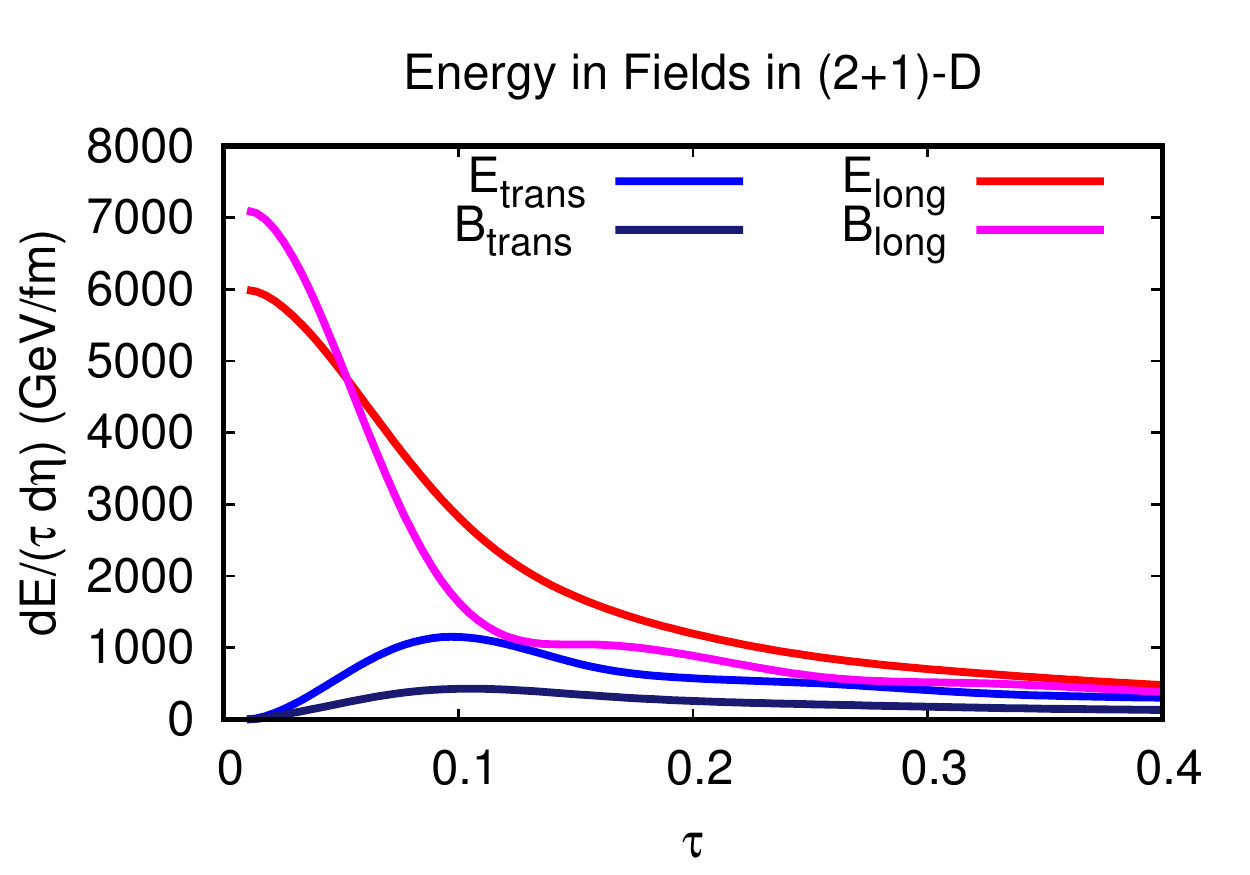}
        \includegraphics[width=0.5\textwidth]{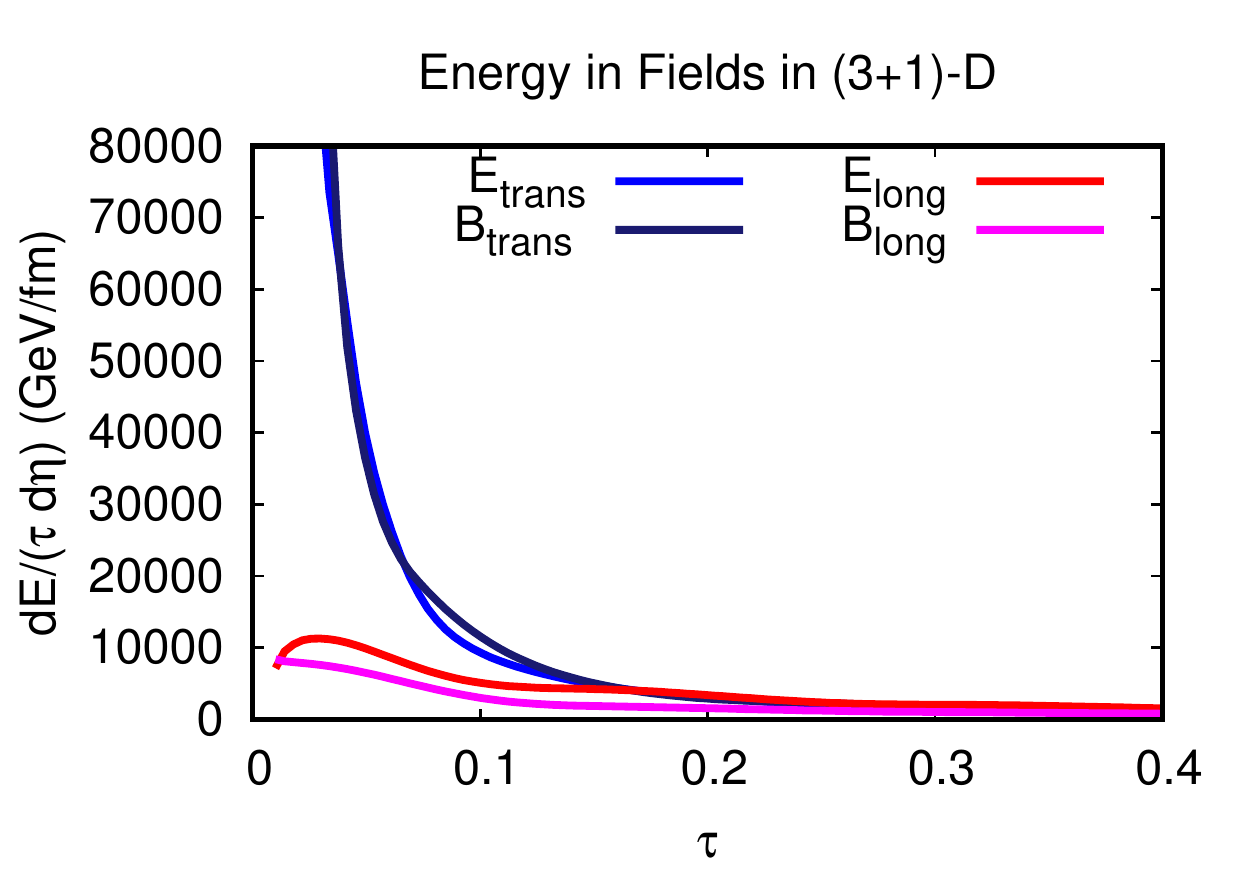}
\caption{Left: The time evolution, in ${\rm fm}$, of the energy density in the
different field components in 2+1D. Right: The same quantity as the left
panel plotted for the 3+1D implementation. Both results are computed using
the same 3+1D software, but with the initial 2+1D and 3+1D setups,
respectively.}
\label{fig1}
\end{figure}

The evolution of the energy density in the fields can be seen for both the
2+1D and 3+1D scenarios in Fig.~\ref{fig1} where we plot
\begin{align}
{1\over \tau}{dE(\tau)\over d\eta} 
& = \int d^2x_\perp 
\left(
\epsilon_\eta + \epsilon_x + \epsilon_y
\right)
\label{eq:Etau}
\end{align}
It is clear that the early
time behaviour is quite different: In 2+1D the transverse fields vanish at
$\tau=0$ and grow steadily until their contribution to the energy density
is comparable to the longitudinal fields, whereas in 3+1D the transverse
fields provide the dominant contribution to the energy density initially.
By typical hydrodynamic initialization times of $\tau=0.2-0.6 \, {\rm fm}$ ,
the 3+1D fields all have similar contributions to the energy, as is the
case in 2+1D.

\begin{figure}[h!]
\centering
       \includegraphics[width=.5\textwidth]{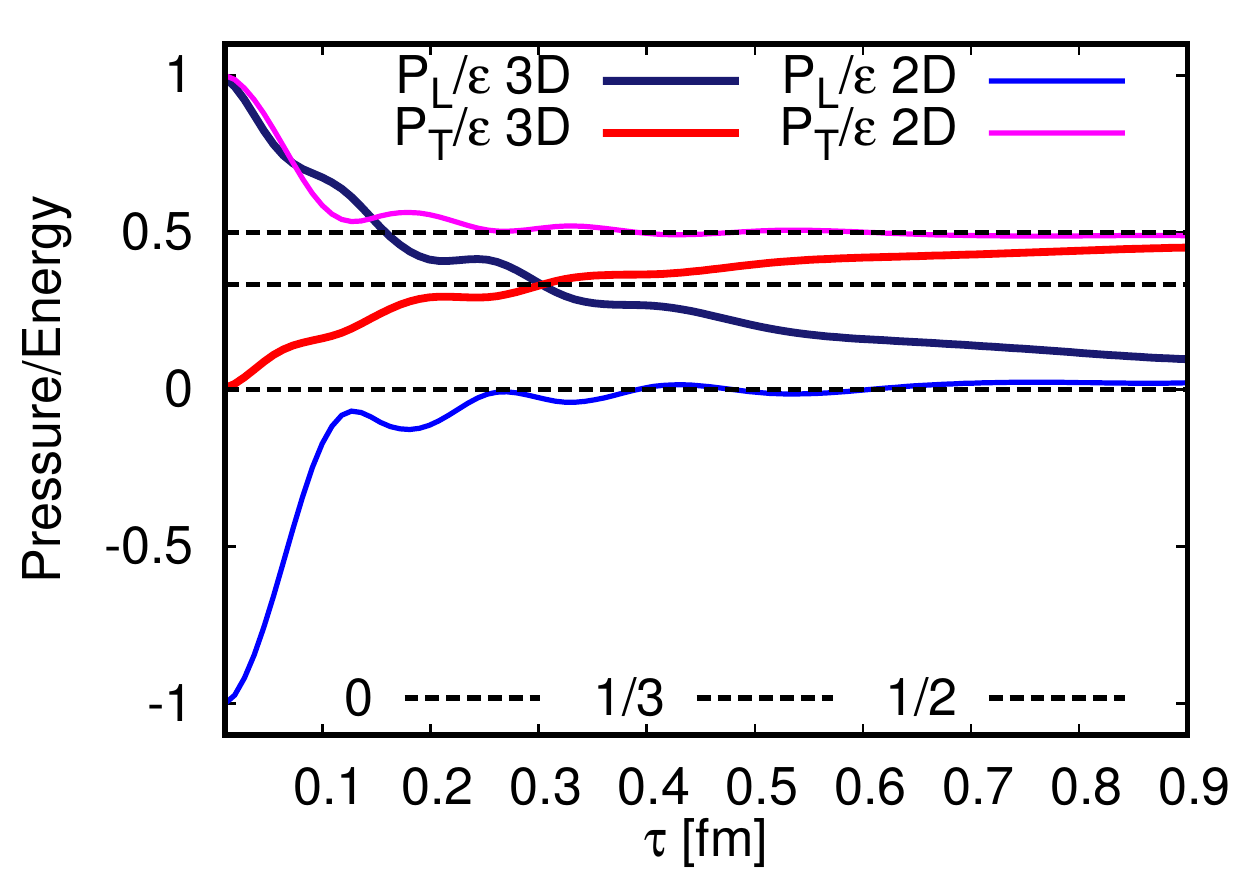}
       \caption{Comparison of the transverse and longitudinal pressures in
the 2+1D and 3+1D IP-Glasma formulations. Both results are computed using
the same 3+1D software, but with the initial 2+1D and 3+1D setups,
respectively.}
       \label{fig:pressure}
\end{figure}

The early time behaviour of the fields in 3+1D causes the longitudinal and
transverse pressures to behave quite differently than in the boost
invariant case
(similar behaviour was observed in Ref.\cite{Gelfand:2016yho}).
To see why, it is convenient to first express the diagonal
components of the stress-energy tensor in terms of the quantities defined
in Eqs.(\ref{eq:epsilon_i}) and (\ref{eq:epsilon_eta}),
\begin{align}
 \begin{split}
    T^{\tau \tau}&= \epsilon_x+\epsilon_y+\epsilon_{\eta}=\epsilon \\
    T^{ii}&=-\, \epsilon_i\,+\epsilon_j+\epsilon_{\eta} \, \biggr
\rvert_{\substack{i=x,y\\j \neq i}} \\
     \tau^2 T^{\eta \eta} &=\epsilon_x+\epsilon_y\, -\,\epsilon_{\eta}.
  \end{split}
\end{align}
where $T^{ii}$ is either $T^{xx}$ or $T^{yy}$.
The pressure to energy ratios are given by,
\begin{align}
      \frac{P_L}{\epsilon} = \frac{\tau^2 T^{\eta\eta}}{T^{\tau \tau}} &&
    \frac{P_T}{\epsilon} = \frac{T^{xx}+T^{yy}}{2T^{\tau \tau}}.
  \end{align}
As can be seen in Fig.~\ref{fig:pressure}, the $\tau \xrightarrow{}0^+$
limit is quite different in 2+1D and 3+1D:
\begin{align}
\lim_{\tau\to 0^+} \frac{P_L}{\epsilon} =
\left\{\begin{array}{ll}
\frac{\epsilon_x+\epsilon_y}{\epsilon_x+ \epsilon_y}=1 & \hbox{in 3+1D}\\
\frac{-\epsilon_\eta}{\epsilon_\eta}=-1 & \hbox{in 2+1D}
\end{array}\right.
\end{align}
and
\begin{align}
    \lim_{\tau\to 0^+} \frac{P_T}{\epsilon} =
     \left\{\begin{array}{ll}
            \frac{\epsilon_\eta}{\epsilon_x+\epsilon_y} = 0 & \hbox{in 3+1D}\\
            \frac{\epsilon_\eta}{\epsilon_\eta}=1 & \hbox{in 2+1D} .
      \end{array} \right.
\end{align}
This because $\epsilon_{x,y}/\epsilon_\eta \sim 1/\tau^2$
in the small $\tau$ limit in 3+1D while $\epsilon_{x,y} = 0$ at $\tau = 0^+$ in 2+1D.

Because of the tracelessness of $T^{\mu\nu}$, the intersection of the pressures
necessarily occurs at $\epsilon/3$ in 3+1D evolution,
the condition for pressure isotropy.
The pressure does not remain isotropic, however, and approaches the 2+1D
asymptotic behaviour for large $\tau$, as the longitudinal pressure
free-streams towards zero in both cases.

\begin{figure}[hb!]
  \centering

\includegraphics[width=0.5\textwidth]{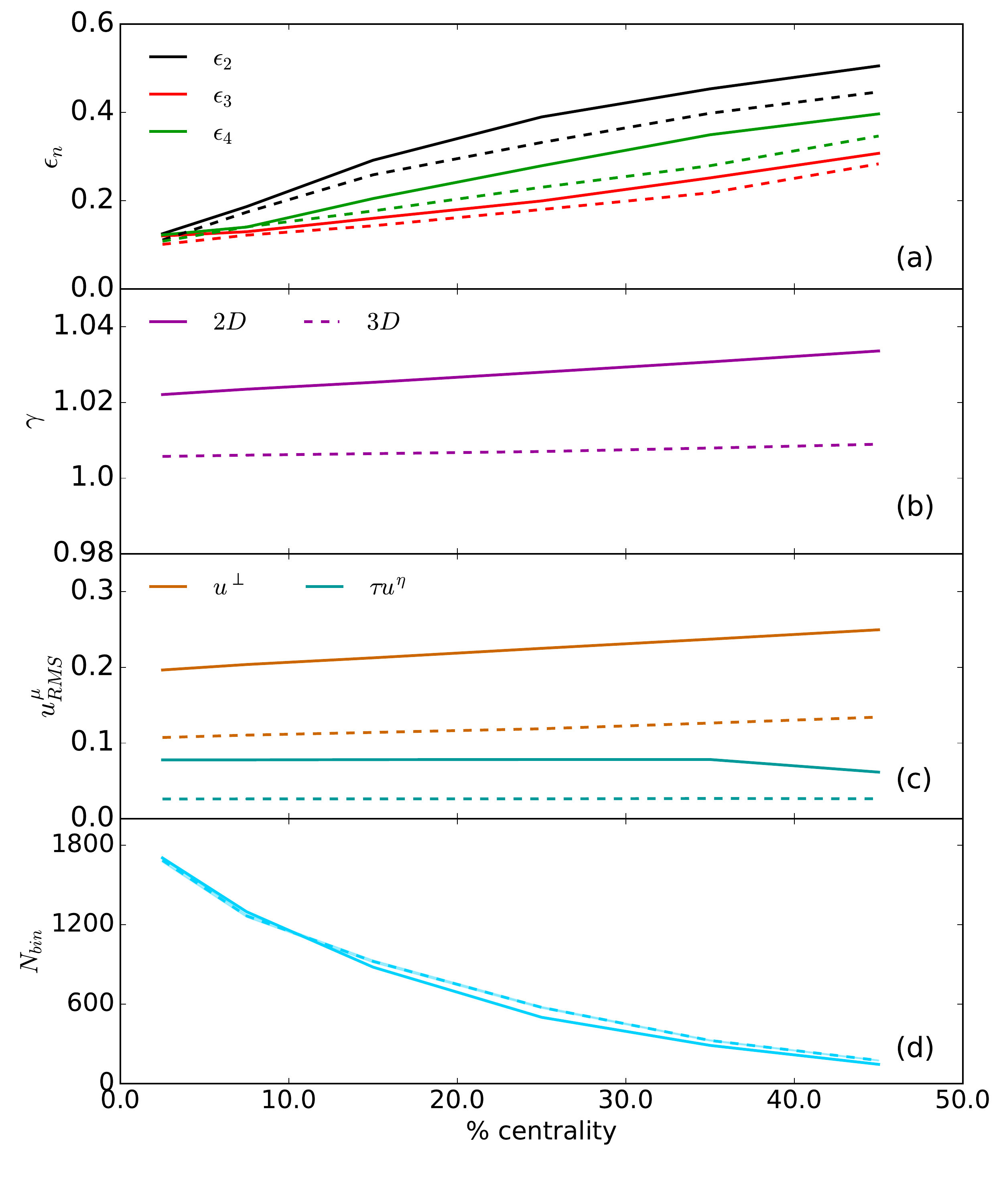}
\caption{A comparison between the 3+1D IP-Glasma presented in this paper
and the 2+1D IP-Glasma in Ref.\cite{McDonald:2016vlt}. Panel (a) compares the
eccentricities between the two simulations. In panel (b), $\gamma=u^\tau$
is plotted, and panel (c) shows the longitudinal and transverse components
of the flow velocity. Finally, panel (d) compares the number of binary
collisions, $N_{bin}$, as a way of gauging how the two different centrality
selection procedures compare. 
The 3+1D curves are calculated at $\tau = 0.6\,\hbox{fm}$
and the 2+1D curves are calculated at $\tau = 0.4\,\hbox{fm}$ which are the respective
hydro switching times.
}
\label{Fig:Init_2D_vs_3D}
\end{figure}

Comparing the 2+1D and 3+1D pressure curves, $P_T/\epsilon$ is
substantially larger in 2+1D for the entire evolution up to the typical hydro
switching time that is used, $\tau=0.4\, {\rm fm}$ This can be seen clearly
in Fig.~\ref{fig:pressure}. More transverse pressure should mean more
transverse flow, and indeed that is what is seen in panel (c) of
Fig.~\ref{Fig:Init_2D_vs_3D},
which compares the transverse and longitudinal
flow between the two simulations. 
One can readily see that the transverse flow develops more rapidly in the 2+1D
simulations.
It needs to be noted that the 3+1D curves
in Fig.~\ref{Fig:Init_2D_vs_3D} correspond to $\tau = 0.6 \, {\rm fm}$,
whereas the 2+1D curves are at $\tau=0.4 \, {\rm fm}$, 
the hydro switching times in the respective simulations. 
One may wonder whether matching the hydronization times
would change any conclusions. Since the flow in the 3+1D case is lower, this is not the case.
Running the 2+1D case up to $\tau = 0.6\,\hbox{fm}$ only accentuates the difference.

\section{Question of the 2+1D Limit}
\label{sec:2DLimit}

In 2+1D, the initial transverse fields $E^i$ and $F_{i\eta}$ are both zero.
This originates from the facts that $A_{0,\eta} = 0$ and that nothing depends on $\eta$.
In 3+1D, the $\eta$ dependence of $E_0^i$ is dictated by the Gauss' law
$[D_i, E_0^i] = -[D_\eta, E_0^\eta]$
and the transverse chromo-magnetic field component
$F_{i\eta}$
depends on $\eta$ through $A_{0,\eta}$ and $A_{0,i}$.
Ultimately, the $\eta$ dependence of any term in these expressions 
comes from the $\eta$ dependence of $V$ in Eq.(\ref{eq:LangevinStep}). 

The main issue for the approach to the 2+1D initial condition,
equivalently the $\sqrt{s}\to\infty$ limit, is how the transverse part of the energy density 
($\tau$ times $\epsilon_{x,y}$ in Eq.(\ref{eq:epsilon_i})) behaves in that limit.
The behaviour of the transverse energy density depends on three main components.
The first one is how fast
the $\eta$ dependence of $V$ goes away as $\sqrt{s}\to\infty$.
The second one is how the initial time $\tau_{0}$ depends on $\sqrt{s}$.
The third one is how fast the field strength grows as $\sqrt{s}$ grows.

The $\sqrt{s}$ dependence of $V$ is in the running coupling constant. 
From Eq.(\ref{eq:zeta_zeta_corr}), one can see that 
$\boldsymbol{\eta} \propto \sqrt{\alpha_s}$
and hence the $\eta$ derivative of $V$ will behave like some power of 
$\alpha_s$. As $\sqrt{s} \to \infty$, the running coupling $\alpha_s \to 0$.
In this sense, one could argue that
the transverse electric field $E^i$ and the magnetic field $F_{i\eta}$ as well as
$A_\eta$ will vanish in the infinite momentum limit, restoring the 2+1D initial conditions.
This argument, however, is too simple. One needs to take into account the behaviour of the gauge field strength and
the behaviour of $\tau_0$ as well.

Consider the transverse magnetic field 
\begin{equation}
F_{i\eta} = \partial_i A_\eta - \partial_\eta A_i - ig[A_i, A_\eta]
\end{equation}
Because of the form of the initial gauge fields given in
Eq.(\ref{eq:MV_Sol_2}) and Eq.(\ref{eq:A_eta_init}), all three terms above contain
an $\eta$-derivative.
The transverse electric field
$E_0^i$ also depends on the size of $\eta$ derivatives through the Gauss' condition.
As argued above, the size of the $\eta$-derivative
is given by some positive power of the strong coupling
$\alpha_s(Q_s)$ where $Q_s$ is the saturation scale.
The transverse gauge field components $A_{0,i}$ behave
like $Q_s/g$ and $\partial_i$ behaves like $Q_s$. 
The initial time $\tau_{0}$ should behave like $1/Q_s$ or $1/\sqrt{s}$.
Putting these all together, one can then argue that at the initial time $\tau_0$
\begin{equation}
\tau_0 \epsilon_{i}
\sim
(F_{i\eta})^2/\tau_0 \sim Q_s^a \alpha_s^b
\end{equation}
where $a, b$ are some positive powers.
Since the saturation scale behaves like a power of $\sqrt{s}$,
($Q^s \sim \sqrt{s}^{\lambda}$) and the coupling constant behaves like
an inverse logarithm of $\sqrt{s}$, ($\alpha_s \sim 1/\ln(Q_s)$),
the transverse energy density $\epsilon_i$ in Eq.(\ref{eq:epsilon_i}) 
cannot vanish as $\sqrt{s} \to \infty$. It will actually diverge.
Hence, our 3+1D initial conditions, although they inherit many features from the 2+1D, will
not recover the 2+1D initial conditions in the infinite momentum limit.

Ultimately, whether or not the boost-invariant initial conditions are recovered 
as $\sqrt{s} \to \infty$
depends on how the $\eta$ derivatives of the initial fields behave
in that limit.
The 2+1D limit will be recovered only if the $\eta$ derivatives vanish faster
than some negative power of $\sqrt{s}$.
There are indeed
other 3+1D Glasma models that do recover the 2+1D limit
\cite{Gelfand:2016yho,Schlichting:2020wrv}.
They can do so because their initial conditions do not include as much
longitudinal fluctuations as those provided by the JIMWLK evolution.
Consequently,
the $\eta$ derivatives in these models go to zero much faster than $\alpha_s$ does.

A related question is whether we should see in our simulations
the Weibel instability observed in various 3+1D CYM simulations
\cite{Romatschke:2005pm,Romatschke:2006nk,Fukushima:2006ax,Fukushima:2011nq,Berges:2012cj,Dusling:2012ig,Epelbaum:2013waa}.
In Ref.\cite{Romatschke:2006nk}, it was shown that the instability-driven 
exponential growth starts around 
$\tau \approx 30/(g^2 \mu)$. Taking $1/(g^2\mu) \sim 1/Q_s = O(0.1)\,\hbox{fm}$ relevant for RHIC and
the LHC collisions \cite{Romatschke:2005pm},
the exponential growth would start around $\tau = O(1)\,\hbox{fm}$.
This is similar or longer than the evolution time of the Glasma in our simulations.
According to this estimate, the effect of such instability would be weak in our simulations.
One should, however,
note that the spectrum of the $\eta$-dependent fluctuations is quite different in those simulations compared to ours.
Investigation of the eventual appearance of the Weibel instability in our setting would be an interesting future study.

\section{Hydrodynamic Evolution and Hadronic Cascade}
\label{sec:hydro_plus_cascade}

After the CYM evolution, the stress energy tensor is constructed from the
chromo-electric and chromo-magnetic fields. The stress-energy tensor is
diagonalized to yield the local energy density, $\epsilon$ and flow
velocity, $u^{\mu}$, via the Landau condition
\begin{equation}
    T^{\mu}_{\,\nu} u^{\nu} = \epsilon u^{\mu}
\end{equation}

In previous studies, it was common to initialize hydrodynamics
simulations with the ideal stress-energy tensor,
\begin{equation}
    T^{\mu \nu}_{{\rm ideal}}= (\epsilon+P)u^{\mu}u^{\nu} - Pg^{\mu\nu}
\end{equation}
In this study, the entire stress-energy tensor that is generated by
3+1D IP-Glasma is used to initialize 3+1D hydrodynamics, and thus no information is
lost in the matching condition.

The hydrodynamic stress-energy tensor can be decomposed into an ideal part
and a viscous part
\begin{equation}\label{eq:fulltmunu}
    T_{{\rm hydro}}^{\mu \nu}
    = T^{\mu \nu}_{{\rm ideal}} +\pi^{\mu \nu} - \Pi (g^{\mu
\nu}-u^{\mu}u^{\nu})
\end{equation}
where $\pi^{\mu\nu}$ is the shear-stress tensor and $\Pi$ is the bulk
pressure. Since the CYM is conformal, there is no bulk pressure, and the
shear stress tensor is simply the difference between the IP-Glasma
energy-momentum tensor and that of ideal hydrodynamics,
\begin{equation}
    \pi^{\mu\nu} = T^{\mu \nu} - T^{\mu \nu}_{{\rm ideal}}
\end{equation}
There is, however, a discontinuity in the relationship between the energy
and pressure, i.e. the equation of state (EoS), in the IP-Glasma phase for
which $\epsilon=3P$ and the hydrodynamic phase for which the EoS comes from
Lattice QCD calculations \cite{Bazavov:2014pvz} matched to a hadronic
resonance gas model. This discontinuity in pressure gives the initial bulk
pressure,
\begin{equation}
    \Pi = {\rm P}_{{\rm CYM}} - {\rm P}_{{\rm hotQCD}}(\epsilon)=
\epsilon/3 - {\rm P}_{{\rm hotQCD}}(\epsilon)
\end{equation}

In this work, the switching time between the CYM dynamics and Hydrodynamics is set to
$\tau_h = 0.6\,\hbox{fm}$.  
This switching time is a little later than the one used in the 2+1D case ($0.4\,\hbox{fm}$)
\cite{McDonald:2016vlt} 
to allow
development of a bit more pre-flow. The effects of changing $\tau_h$ are, 
however, not extensively
studied in the 3+1D setting so far and will be left for future studies.
The values of the transport coefficients we used are the same as 
those we used in Ref.\cite{McDonald:2016vlt} except the value of the shear viscosity which is set
to $\eta/s = 0.08$ due to the fact that hydrodynamic flow develops slower in the 3+1D expansion than 
in the 2+1D expansion as explained in the next section. 
Switch to UrQMD occurs via Cooper-Frye formula at the hypersurface defined by the switching
temperature of $145\,\hbox{MeV}$.

Altogether, 1,200 3+1D IP-Glasma$+$ Music events were generated between in the $0 - 50\,\%$ 
centrality range, or 240 per $10\,\%$ in each $10\,\%$ centrality bin. 
Each one of these events was then sampled for 100 UrQMD runs.

\section{Results}
\label{sec:results}

\subsection{Initial state quantities}

Before describing the 3+1D results, it is important for us
to check whether the additional physics present in
the initialization of the fields leads to any differences in the
mid-rapidity physics. For this purpose we show the initial state anisotropy
(a.k.a.~eccentricity) as characterized by
\begin{equation}
    \varepsilon_n = \frac{\int{d^2x_\perp r^n
e^{in\phi}\epsilon({\bf x}_\perp)}}{\int{d^2x_\perp r^n\epsilon({\bf x}_\perp)}},
\end{equation}
Here $\epsilon({\bf x}_\perp)$ is the energy density at ${\bf x}_\perp$, 
$r = |{\bf x}_\perp|$ and $\phi = \tan^{-1}(y/x)$.
Panel (a) of Fig.~\ref{Fig:Init_2D_vs_3D} compares the eccentricities
between the 3+1D and 2+1D simulations, where the 2+1D simulations are from
Ref.\cite{McDonald:2016vlt}. The two simulations show the same trends but the
$\varepsilon_n$ values are systematically larger in the 2+1D simulation,
particularly at larger centralities. This could be partially due to 
the smaller number of events for the 3+1D case.
The 3+1D events have an
order of magnitude fewer events than the 2+1D. As such, it is possible that the
tail of the 3+1D multiplicity distribution was not fully populated.
Panel (d) of  Fig.~\ref{Fig:Init_2D_vs_3D} compares the number of binary
collision, $N_{bin}$, as a function of centrality, as a way of showing how
nucleus-nucleus overlap corresponds to centrality. This figure shows that the
2+1D curve is slightly steeper as a function of centrality, which is
consistent with the more rapid rise in $\varepsilon_n$ as a function of
centrality shown in panel (a).  The centrality selection done here follows
the same procedure discussed in Ref.\cite{McDonald:2016vlt}.

In panel (b) and (c), the flow vector components $u^\tau = \gamma$, and the RMS values of
the spatial components defined as
\begin{equation}
u^\mu_{\rm RMS} = \sqrt{ \langle
{
\int d^2x_\perp \epsilon({\bf x}_\perp)(u^\mu)^2
\over
\int d^2x_\perp \epsilon({\bf x}_\perp)
}
\rangle }
\end{equation}
are shown. The angular bracket here means the average over the events.
These values are measured at the hydro-switching time $\tau = 0.6\,\hbox{fm}$ 
for the 3+1D simulations
and $\tau = 0.4\,\hbox{fm}$ for the 2+1D simulations. 
Although the 3+1D CYM
simulations were allowed to run longer, the flow components are still smaller than
the 2+1D case.  
This ultimately results from the difference in the behaviours of the pressure component
which was discussed in depth in section \ref{sec:Pressure}.
One consequence of less-developed pre-flow is that the value of
shear viscosity over entropy density, $\eta/s$, in 3+1D simulations needs to be less
than that in 2+1D
simulations to match the experimental data.

\subsection{Mid-rapidity Observables}

While the hydrodynamic evolution allows for some tuning of parameters, such
as the transport coefficients, the 3+1D IP-Glasma initialization is able
to describe the essential mid-rapidity observables. 
Here, $\eta/s=0.08$, which differs from the value favoured by 2+1D IP-Glasma.
For example, Ref.\cite{Schenke:2020mbo} which is a recent 2+1D IP-Glasma paper 
and uses the same EOS as this work has $\eta/s = 0.12$. Our previous calculations using 2+1D IP-Glasma
\cite{McDonald:2016vlt} used $\eta/s = 0.095$. The difference in $\eta/s$ is
due in part to the differences in pre-equilibrium flow between the 2+1D and
3+1D simulations. The bulk viscosity, $\zeta/s(T)$, is taken from
Ref.\cite{Ryu:2015vwa} and is consistent with that typically used in 2+1D
IP-Glasma simulations, and finds similar agreement with data.

\begin{figure}[hb!]
  \centering
\includegraphics[width=0.4\textwidth]{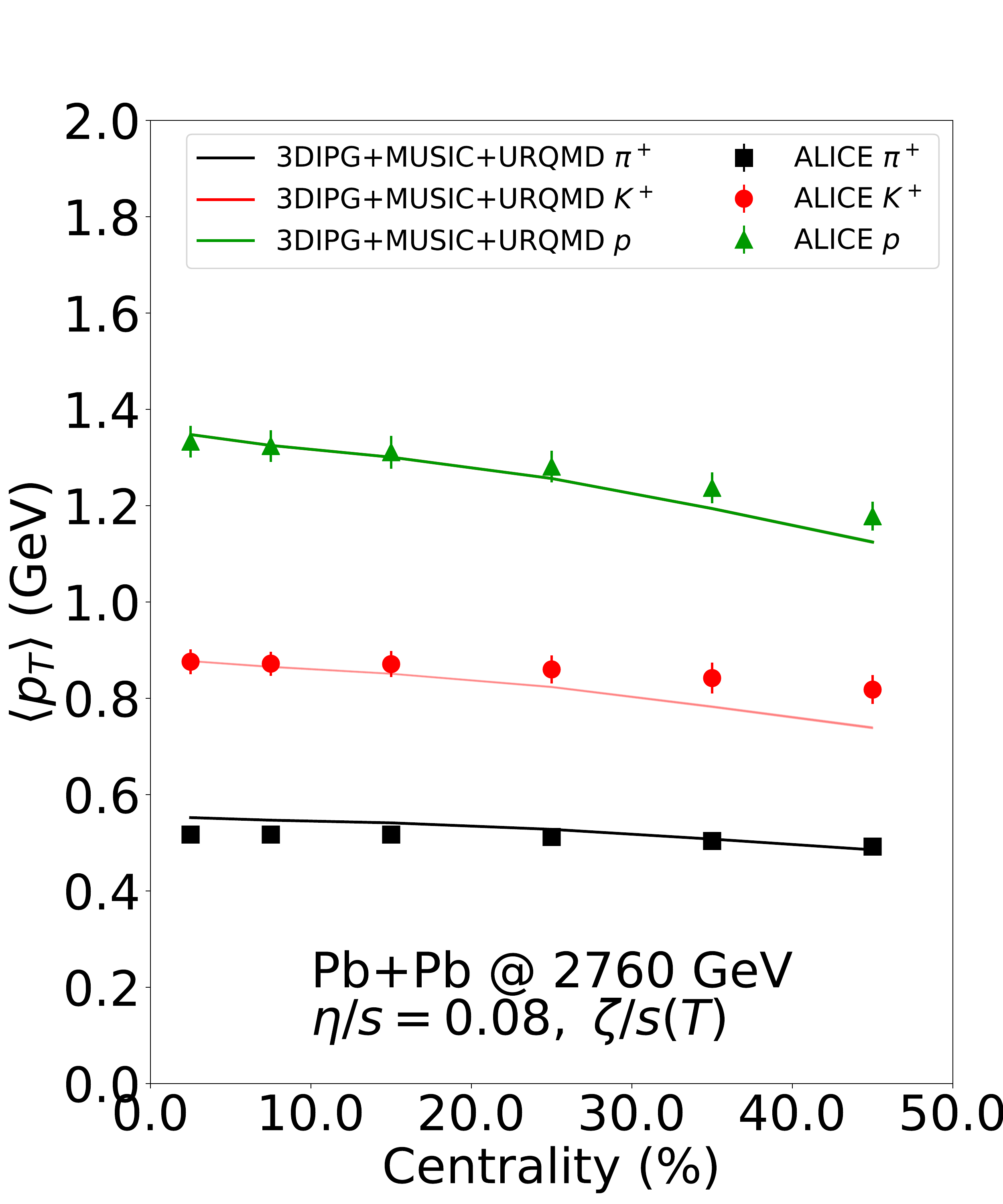}
\caption{The particle identified $\langle p_T \rangle$ for Pb-Pb at 2.76
TeV in the 0-5\% centrality bin as compared to ALICE data
\cite{Abelev:2013vea}.}
    \label{fig:meanpt}
\end{figure}

\begin{figure*}
    \centering
    \includegraphics[width=1.0\textwidth]{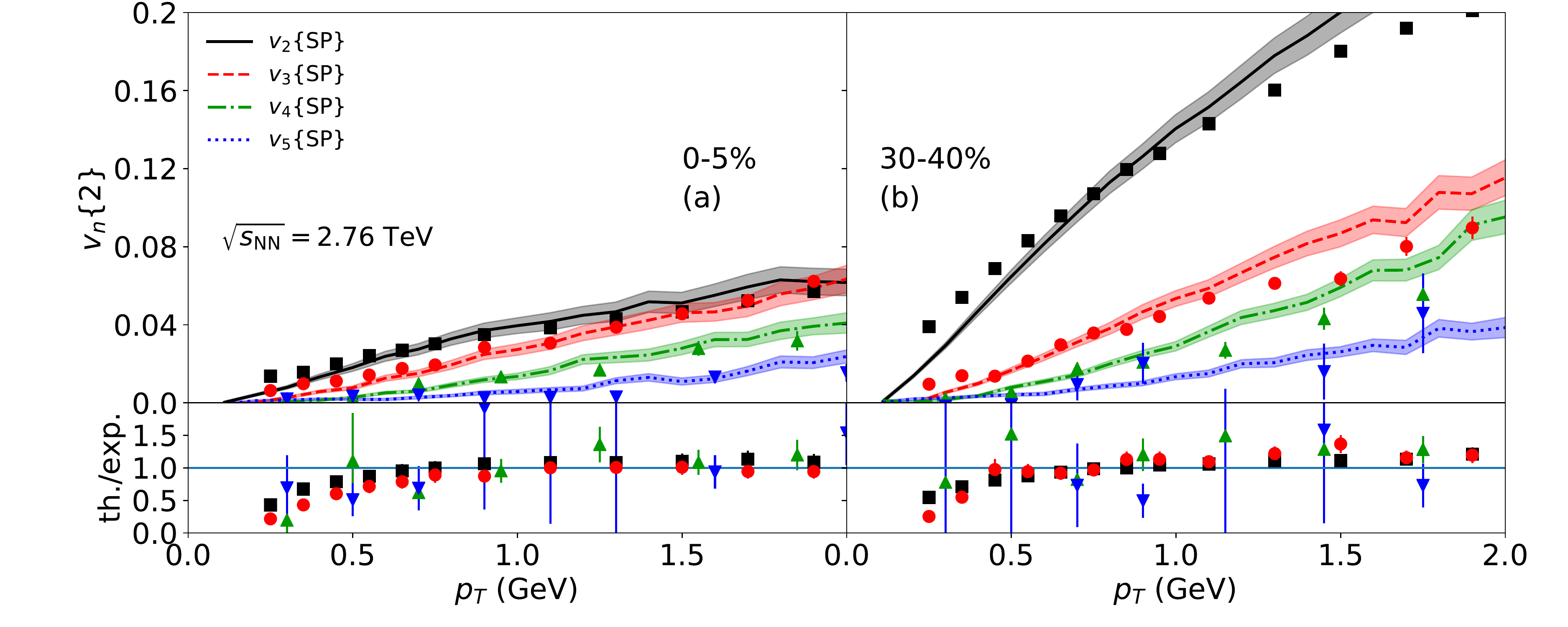}
    \caption{Differential flow harmonics $v_n\{2\}(p_T)$ for two
centralities, compared to ALICE data \cite{ALICE:2011ab}. Left upper panel
is for $0-5\%$ and right upper panel is $30-40\%$ centrality. Lower panels
show the ratio of theoretical data to experimental data from the upper
panels.}
\label{fig:vn_spectra}
  \vspace*{\floatsep}

\includegraphics[width=1.0\textwidth]{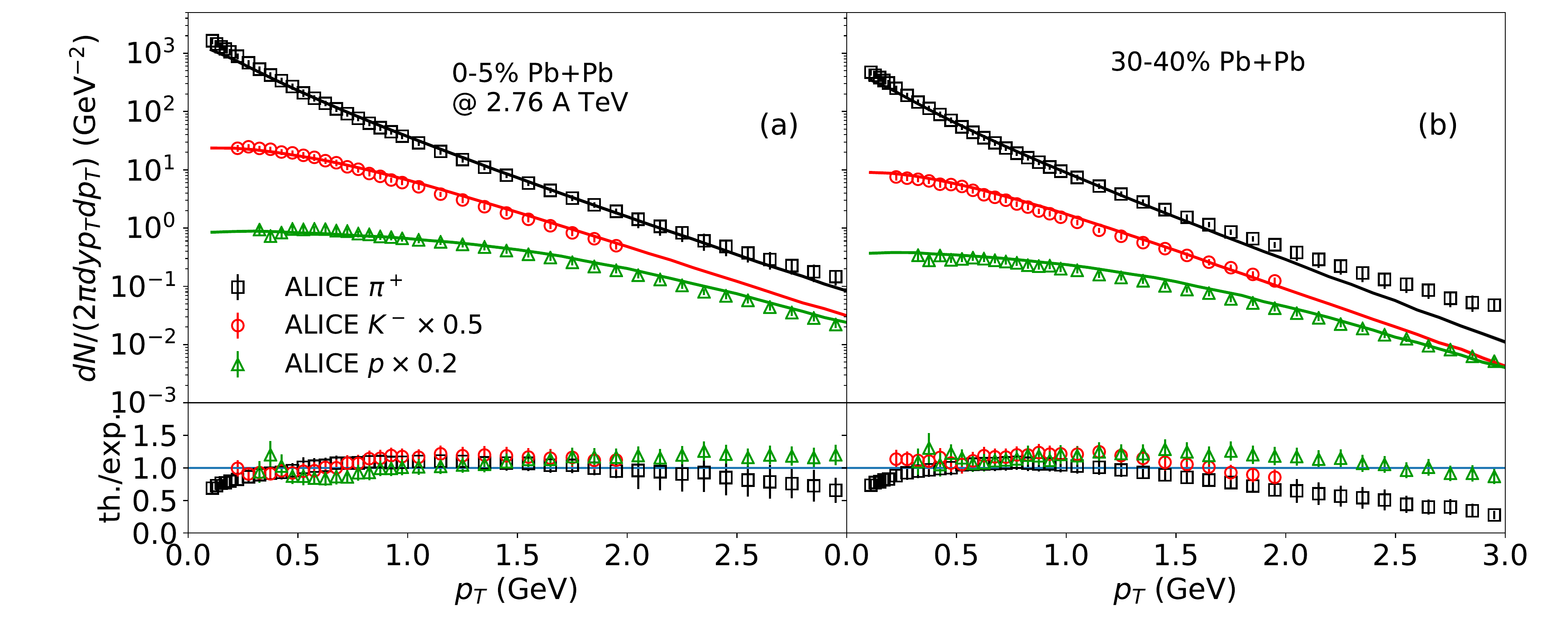}
    \caption{Identified particle spectrum for two centralities, compared to
ALICE data \cite{Abelev:2013vea}. The lower panels show the ratio of the
theoretical data to the experimental calculation for each curve.}
    \label{fig:spectra}
\end{figure*}

The hadronic spectra is well-described, shown for two centrality classes in
Fig.~\ref{fig:spectra}, as are the particle identified $\langle p_T
\rangle$ in Fig.~\ref{fig:meanpt}. The differential $v_n$'s 
shown in Fig.~\ref{fig:vn_spectra} have similar
behaviour to other hydrodynamic calculations that include bulk viscosity: a slight
underestimate of the $v_n(p_T)$ at small $p_T$, say up to $p_T \approx 0.7
\, {\rm GeV}$ and a slight overestimate above. Hydrodynamic calculations without
bulk viscosity are typically able to describe the $v_n(p_T)$ over a much
wider range in $p_T$ but typically overestimate the $\langle p_T \rangle$,
particularly for heavier particles such as protons. 
By including bulk
viscosity, one typically improves the spectra and $\langle p_T \rangle$,
but degrades agreement with $v_n(p_T)$. It is still possible to find good
agreement with the integrated $v_n$, by missing the $v_n(p_T)$ at small
$p_T$ and missing in the opposite direction at high $p_T$. 
In this study, the specific shear viscosity $\eta/s$ is tuned to find agreement with the
integrated $v_2$. 

To summarize, phenomenologically, including bulk
viscosity leads to the trade off of $v_n(p_T)$ for $\langle p_T \rangle$
and the particle spectra. This can be justified by the fact the spectra and
$\langle p_T \rangle$ are more basic quantities and that $v_n(p_T)$
is sensitive to $\delta f$ corrections that have large uncertainties,
particularly in the case of bulk viscosity.

\subsection{Rapidity Dependent Observables}
\begin{figure}[hb!]
  \centering

\includegraphics[width=0.5\textwidth]{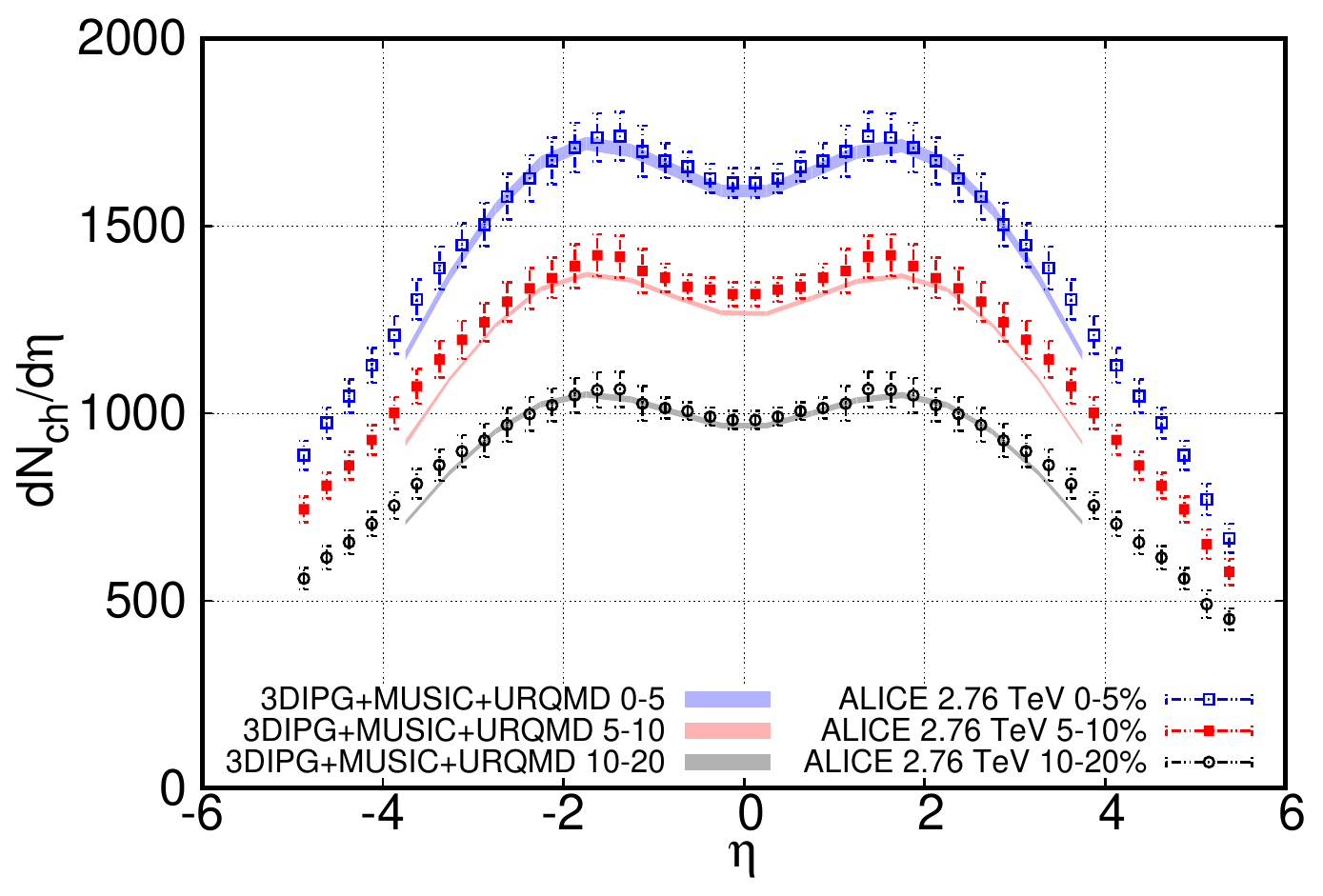}
\caption{Charged hadron multiplicity as a function of rapidity, as compared
to ALICE data\cite{Abbas:2013bpa}.}
\label{dndeta}
\end{figure}
The primary purpose of developing a 3-dimensional extension of the boost
invariant IP-Glasma is to explore the longitudinal dynamics of HIC's, and
we do so in this section.

The initial state events are run in a rapidity window of [-4,4].  In order
to avoid sharp gradients at the boundaries in $\eta$, it is necessary to
put an envelope function on the hydrodynamic evolution that provides a
smooth gradient to zero density for $|\eta|>2.75$. Here, a half Gaussian
takes the components of $T^{\mu\nu}$ to zero as follows

\begin{align}
 \begin{split}\label{eq:eta_profile}
    T^{\mu\nu}_{{\rm hydro}}(\mathbf{x_\perp}, \eta, \tau_h) &=
T^{\mu\nu}_{\hbox{\scriptsize IP-Glasma}}(\mathbf{x_\perp}, \eta, \tau_h)\\
&\exp{\left[-\theta\left(|\eta|-2.75|\right)\frac{(|\eta|-2.75)^2}{2}\right]}.
\end{split}
\end{align}
With this envelope, the model can match the pseudo-rapidity dependence of
the charged hadron multiplicity per unit pseudorapidity, $dN_{ch}/d\eta_s$,
as plotted in Fig.~\ref{dndeta}.

\begin{figure*}
    \centering
\includegraphics[width=1.0\textwidth]{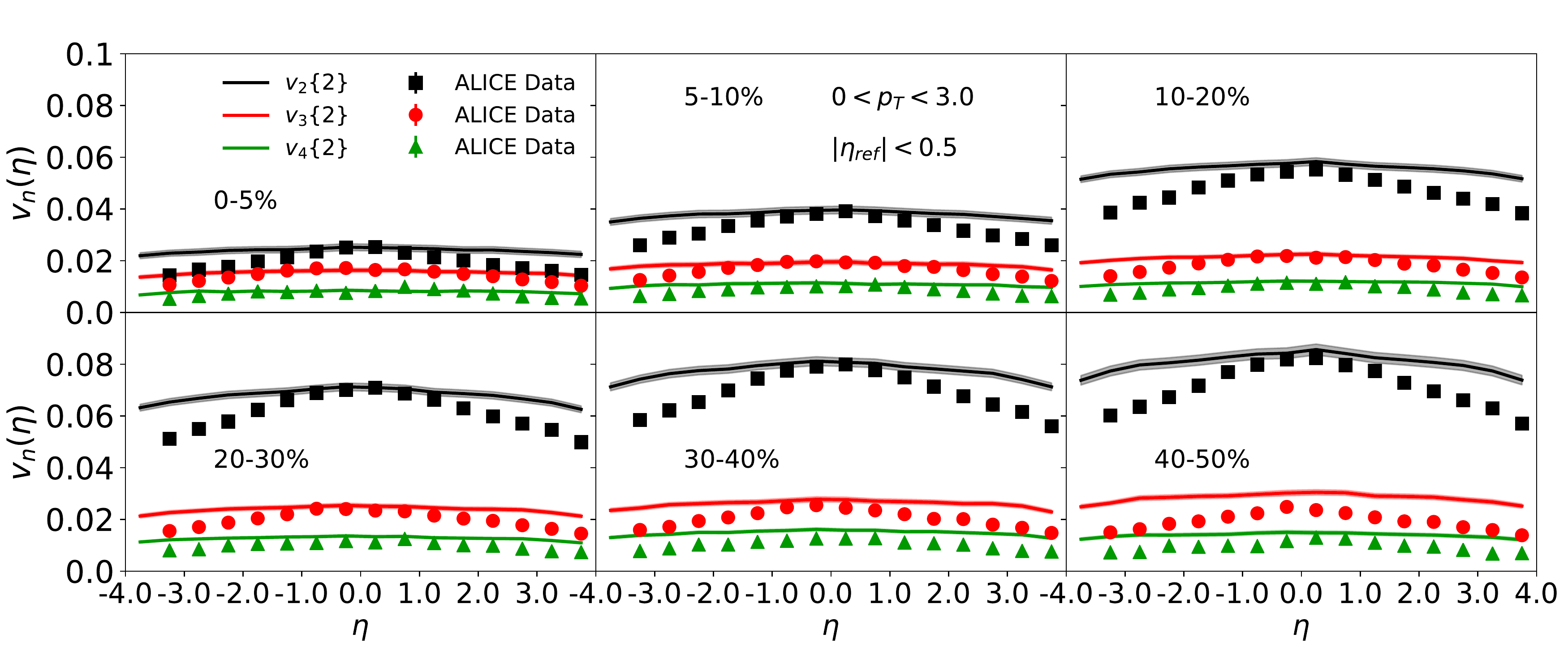}
    \caption{ The rapidity dependence of the momentum anisotropies
$v_n(\eta) (n=2,3,4)$, compared to ALICE data \cite{Adam:2016ows}. Both
data and the calculation are for $p_T>0\, GeV$ and use reference particles
at mid-rapidity ($\lvert \eta \rvert \le 0.5$).}
    \label{fig:vn_eta_ALICE}
  \vspace*{\floatsep}
    \includegraphics[width=1.0\textwidth]{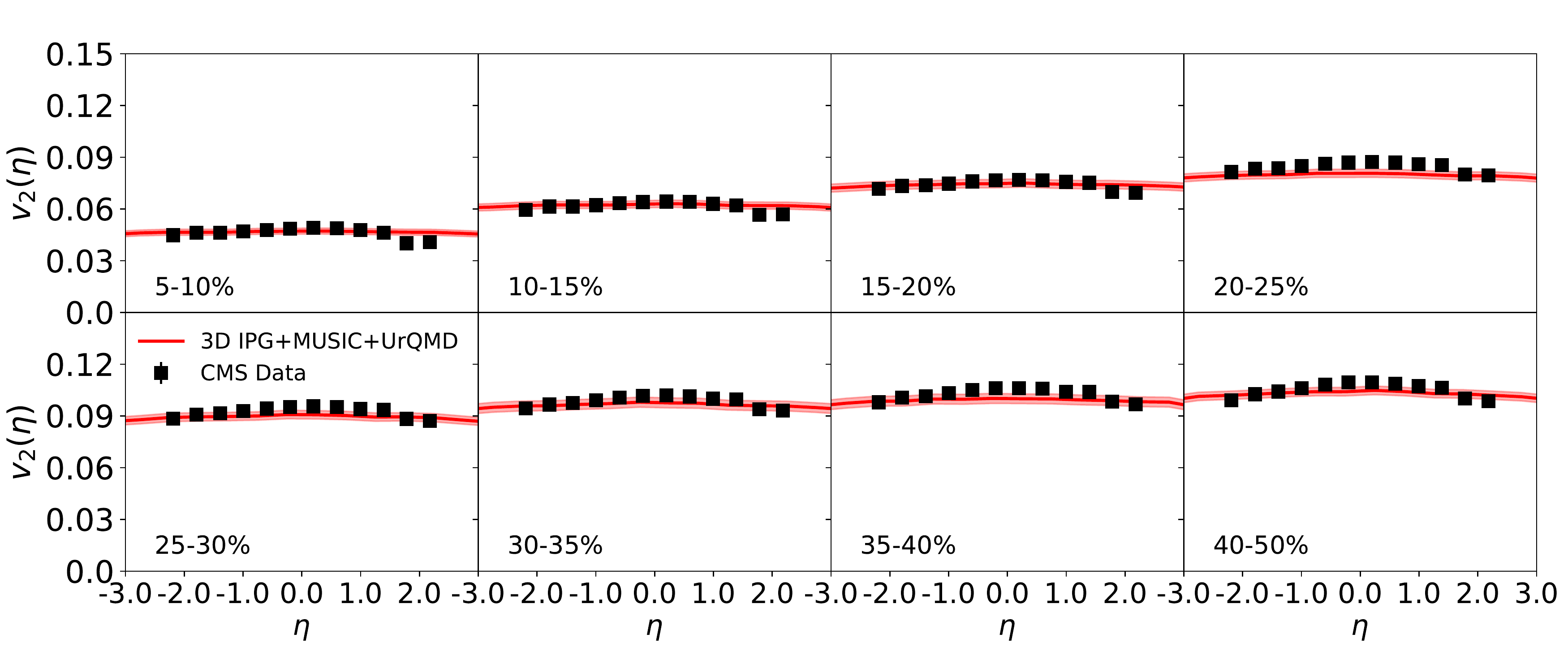}
    \caption{The rapidity dependence of $v_2(\eta)$, compared to CMS data
\cite{Chatrchyan:2012ta}. Both the CMS data and the calculation are for
$0.3 <p_T < 3.0 \, {\rm GeV}$ and references particles in $\lvert \eta
\rvert \le 2.4$.}
    \label{fig:vn_eta_CMS}
\end{figure*}

The multiplicity distribution $dN_{ch}/d\eta_s$ does not 
really require sophisticated IP-Glasma initial states to describe 
\cite{Schenke:2010nt}
as it is not very sensitive to longitudinal fluctuations. 
To see the effect of longitudinal fluctuations better, correlation observables are needed.
In Fig.~\ref{fig:vn_eta_ALICE} and Fig.~\ref{fig:vn_eta_CMS}, we show our calculations of 
the flow harmonics as a function of the pseudorapidity
following the procedures in Ref.\cite{Adam:2016ows} (by ALICE) 
and Ref.\cite{Chatrchyan:2012ta} (by CMS).
The main differences between the ALICE and the CMS measurements are the reference ranges 
($|\eta| < 0.5$ for ALICE and $|\eta| < 2.4$ for CMS) and the $p_T$ ranges
($p_T > 0$ for ALICE and $0.3\,\hbox{GeV} < p_T < 3.0\,\hbox{GeV}$ for CMS).

In Fig.~\ref{fig:vn_eta_CMS}, $v_2(\eta)$ is compared to CMS data, using
their kinematic cuts of $p_T$ and $\eta$. The flow harmonic $v_2(\eta)$ has
a very mild rapidity dependence and the calculation shows similar behaviour.
The CMS data uses reference particles over a range $\eta$
that is a much wider range than that 
used by ALICE in Fig.~\ref{fig:vn_eta_ALICE}. 
This likely contributes to the
the steeper rapidity dependence in the ALICE data, because one would expect
a more peaked structure at mid-rapidity when correlating with mid-rapidity,
as seen in the data. It is, however, apparent that the current 3+1D IP-Glasma 
initial conditions do not contain enough longitudinal decorrelations 
to describe the ALICE data. This may be remedied by introducing thermal fluctuations
in the hydrodynamic evolution. 
It is also possible that the fact our calculations underestimates 
$v_n(p_T)$ in the low momentum region may also contribute to the discrepancy,
but this needs to be investigated further.

We have also calculated the rapidity correlation $r_n(\eta_a, \eta_b)$ \cite{McDonald:2020oyf}.
However, as the number of events we have so far
(240 3+1D-IP-Glasma$+$MUSIC events per 10\,\% centrality) turned out to be too small to make
statistically meaningful statements, we will leave it for future study.

\section{Conclusion}
\label{sec:conclusion}

The purpose of this study is to introduce a realistic model of 3+1D initial
conditions for relativistic heavy ion collision simulations.
The IP-Glasma model, originally developed for 2+1D simulations, has had
great phenomenological success for description of the mid-rapidity observables
that reflect the underlying QGP dynamics.
To extend the reach of theoretical descriptive and predictive power to 3+1D, 
it is imperative to develop realistic extension of the IP-Glasma initial conditions. 
Furthermore, hydrodynamic and hadronic cascade simulations of heavy ion
collisions are capable of handling 3+1D dynamics.
As these simulations are sensitive to the
initial conditions, it is crucial to develop realistic 3+1D initial
conditions. 

Owing to the fact that the simplicity of the 2+1D formulation comes from the assumption 
of the infinite momentum frame (equivalently, boost invariance), 
the extension is not just a matter of trivially adding one more dimension to the 2+1D IP-Glasma.
In this study, we have made an effort to preserve the simplicity 
of the 2+1D formulation as much as possible while breaking the boost invariance.

Our way of doing so is to generate longitudinal structure in the
pre-collision gluon fields through the JIMWLK evolution, the numerical
implementation of which was developed in Ref.\cite{Lappi:2012vw}. This was
incorporated in the IP-Glasma model in Ref.\cite{Schenke:2016ksl}. There
remains theoretical difficulties in temporally evolving this system on the
lattice in three spatial dimensions, however. These include the difficulty
posed by the initial gauge fields and the initial solution to Gauss' law,
as outlined in Section \ref{sec:3D_init_gauge_field}. Both of these problems are
addressed in this work, although the solutions may not be unique. This allows for
a temporal evolution in three spatial dimensions and thus exploration of
the the phenomenological effects of the longitudinal structure generated by
the JIMWLK equations. The 3+1D IP-Glasma simulation is coupled to MUSIC and
UrQMD for comparison to hadronic results.

The 2+1D IP-Glasma describes the transverse dynamics of heavy ion collisions
extremely well. With slightly modified parameters, the 3+1D implementation is
able to achieve similar level of
agreement to key observables such as $\langle p_T
\rangle$, particle spectra, and $p_T$-integrated $v_n$.
In addition, the 3+1D IP-Glasma is able to explore longitudinal observables.
In this paper, the multiplicity and $v_n$ flow harmonics are explored as a
function of pseudo-rapidity, and good agreement is found. Comparison to
higher order correlations involving the longitudinal direction will be
explored in a future work, once substantially better statistics are
generated. This work serves as a proof of principle that the IP-Glasma can
be generalized to 3+1D in a way that allows for consistent temporal
evolution on the lattice and thus phenomenological application.

\begin{acknowledgments}
We acknowledge the support of the Natural Sciences and Engineering Research Council of Canada (NSERC), 
[SAPIN-2018-00024 and SAPIN-2020-00048].
Computations were made on the Beluga supercomputer system from McGill University,
managed by Calcul Qu\'ebec (calculquebec.ca)
and Digital Research Alliance of Canada (alliancecan.ca).
The operation of this supercomputer is funded by the Canada Foundation for Innovation (CFI), 
Minist\`ere de l'\'Economie, des Sciences et de l'Innovation du Qu\'ebec (MESI) 
and le Fonds de recherche du Qu\'ebec -- Nature et technologies (FRQ-NT).

We also gratefully acknowledge
R.~Fries, A.~Ipp, B.~Schenke, S.~Schlichting and R.~Venugopalan
for insightful discussions.

\end{acknowledgments}

\appendix
\section{Solution to Gauss' Law}\label{appendix:gauss_law_numerical}

As mentioned, using the ansatz in Eq.(\ref{eq:phi_ansatz}), turns Gauss'
law Eq.(\ref{eq:gauss}) into the covariant Poisson equation.  We use a
modified Jacobi method for solving the Poisson equation to find the initial
transverse E-fields that satisfy Gauss' Law:
\begin{align*}
    \nabla_\perp^2\phi = -\rho
\end{align*}
Discretizing, and solving for $\phi_{i,\,j}$:
\begin{align}
    \frac{\phi_{i+1,\,j}+\phi_{i-1,\, j}-2\phi_{i,
\,j}}{h^2}+\frac{\phi_{i, \,j+1} + \phi_{i,\, j-1}-2\phi_{i,\,
j}}{h^2}=-\rho_{i,\, j}
\end{align}
\begin{align}
    \phi_{i,\,j}=\frac{1}{4}(\phi_{i+1,\,j}+\phi_{i-1,\,j} +\phi_{i,\,j+1}
+\phi_{i,\,j-1}+h^2\rho_{i,\,j})
\end{align}

Then the iterative procedure is given by:

\begin{align}
    \phi_{i,\,j}^{n+1}=\frac{1}{4}(\phi_{i+1,\,j}^n+\phi_{i-1,\,j}^n
+\phi_{i,\,j+1}^n +\phi_{i,\,j-1}^n+h^2\rho_{i,\,j}^n)
\end{align}
For covariant derivatives, all quantities should be parallel transported:
\begin{align}
\phi_{i,\,j}^{n+1}&=\frac{1}{4}(U_{i,\,j}\phi_{i+1,\,j}^nU^\dagger_{i,\,j}
+U^\dagger_{i-1,\,j}\phi_{i-1,\,j}^nU_{i-1, \, j} + \\ &U_{i,\,
j}\phi_{i,\,j+1}^nU^\dagger_{i,\,j}
+U^\dagger_{i,\,j-1}\phi_{i,\,j-1}^nU_{i,\, j-1}+h^2\rho_{i,\,j}^n)
\end{align}

\section{Conservation of Energy}

\begin{figure}[b]
    \centering
    \includegraphics[width=1.0\linewidth]{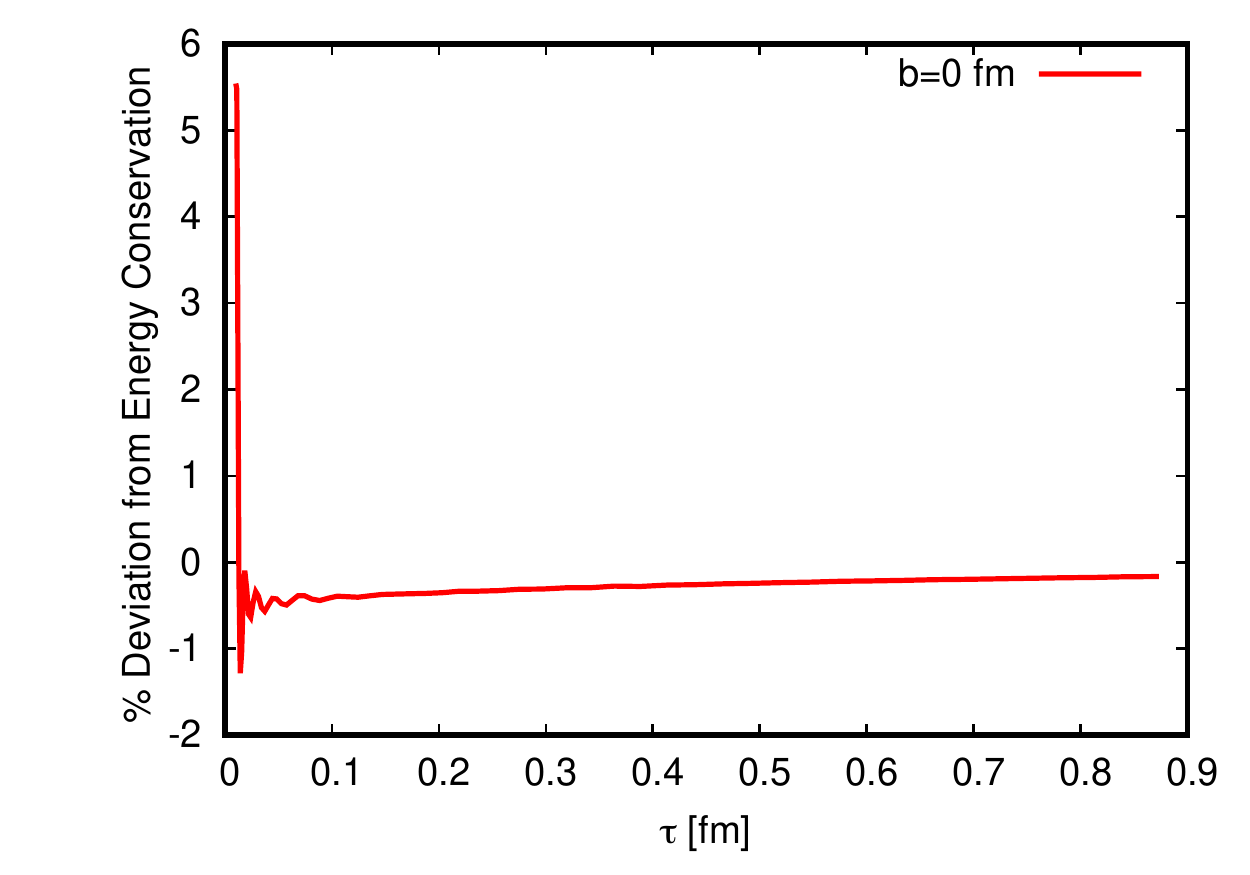}
    \caption{Plotted is the temporal evolution of
\ref{eq:energy_cons_ratio}. Despite fluctuations at very early times,
energy is nicely conserved throughout the evolution of the system. This
figure is a typical event with $b=0 \, {\rm fm}$.}
    \label{fig:energy_conservation_plot}
\end{figure}

The statement of energy conservation in Milne-coordinates is
\begin{equation}
    \partial_\tau T^{\tau \tau} + \partial_\perp  T^{\perp \tau} +
\partial_\eta  T^{\eta \tau} =  -\tau T^{\eta\eta} -T^{\tau\tau}/\tau .
\end{equation}
Multiplying by $\tau$ and collecting terms gives
\begin{equation}
    \partial_\tau (\tau T^{\tau \tau}) + \partial_\perp (\tau T^{\perp
\tau}) + \partial_\eta (\tau T^{\eta \tau}) =  -\tau^2 T^{\eta\eta} .
\end{equation}
Integrating over the 4-dimensional volume ($dx dy d\eta d\tau$)
\begin{equation}
    (\Delta T^{\tau \tau}_{total}) + \int{dx dy d\eta d\tau
\partial_\eta(\tau T^{\eta\tau})}= -\int{dx dy d\eta d\tau (\tau^2
T^{\eta\eta})} .
\end{equation}
The second term becomes a boundary term,
\begin{equation}\label{energy_cons_eq}
    (\Delta T^{\tau \tau}_{total}) + \left( \int{dx dy d\tau (\tau
T^{\eta\tau})}\right)\biggr\rvert_{\eta_{min}}^{\eta_{max}}= -\int{dx dy
d\eta d\tau (\tau^2 T^{\eta\eta})}
\end{equation}

In Fig.~\ref{fig:energy_conservation_plot}, the deviation of the ratio of
the LHS to the RHS of Eq. (\ref{energy_cons_eq}) from unity is plotted.
Explicitly, the quantity on the y-axis is
\begin{equation} \label{eq:energy_cons_ratio}
   \left( 1- \frac{\int{dx dy d\eta (\tau T^{\tau
\tau})}\biggr\rvert_{\tau_{min}}^{\tau_{max}} + \left( \int{dx dy d\tau
(\tau T^{\eta\tau})}\right)\biggr\rvert_{\eta_{min}}^{\eta_{max}}}{
-\int{dx dy d\eta d\tau (\tau^2 T^{\eta\eta})} } \right)\times 100\%.
\end{equation}
There is deviation from energy conservation at extremely early times,
likely due to lattice effects, but the ratio approaches and remains close
to zero for the rest of the evolution. This indicates that energy is
conserved to within about one percent throughout most of the simulation.

\end{document}